\documentclass[prl,floatfix,showpacs,twocolumn,preprintnumbers,amsmath,amssymb,superscriptaddress]{revtex4}
\usepackage{graphicx,float,datetime}
\usepackage[ansinew]{inputenc}
\usepackage{color}
\usepackage{ragged2e}
\usepackage{scalerel}
\usepackage{booktabs}
\usepackage[toc,page]{appendix}
\usepackage{textpos}
\usepackage{ctable}
\usepackage{siunitx}
\usepackage[justification=justified]{caption}

\newcommand{\udt}[3]{#1^{#2}_{\phantom{#2}#3}}

\newcommand{\dut}[3]{#1_{#2}^{\phantom{#2}#3}}

\newdateformat{mydate}{\THEDAY\hspace{3pt}\monthname[\THEMONTH] \THEYEAR}

\begin{document}

\begin{center}
\title{Galactic Rotation Dynamics in $f(T)$ gravity}
\date{\mydate\today}
\author{Andrew Finch\footnote{andrew.finch.12@um.edu.mt}}
\affiliation{Institute of Space Sciences and Astronomy, University of Malta, Msida, MSD 2080, Malta}
\affiliation{Department of Physics, University of Malta, Msida, MSD 2080, Malta}
\author{Jackson Levi Said\footnote{jackson.said@um.edu.mt}}
\affiliation{Institute of Space Sciences and Astronomy, University of Malta, Msida, MSD 2080, Malta}
\affiliation{Department of Physics, University of Malta, Msida, MSD 2080, Malta}

\begin{abstract}
We investigate galactic rotation curves in $f(T)$ gravity, where $T$ represents a torsional quantity. Our study centers on the particular Lagrangian $f(T)=T+\alpha{T^n}$, where $|n|\neq 1$ and $\alpha$ is a small unknown constant. To do this we treat galactic rotation curves as being composed from two distinct features of galaxies, namely the disk and the bulge. This process is carried out for several values of the index $n$. The resulting curve is then compared with Milky Way profile data to constrain the value of the index $n$ while fitting for the parameter $\alpha$. These values are then further tested on three other galaxies with different morphologies. On the galactic scale we find that $f(T)$ gravity departs from standard Newtonian theory in an important way. For a small range of values of $n$ we find good agreement with data without the need for exotic matter components to be introduced.
\end{abstract} 


\maketitle

\end{center}

\section{I. Introduction}
Over the past several decades, the consistent missing mass or dark matter problem has attracted increasing interest in modified and alternative theories of gravity. This is one of the biggest potential contentions between general relativity (GR) and observational astronomy \cite{sanders2010dark}. In fact, in essentially every observed galaxy, it appears that the expected rotational velocities of Newtonian gravity do not conform with the measured values. Since the discrepancy was first noted, observational techniques have drastically improved, making it possible to study and constrain the motions of luminous matter with much greater precision \cite{1538-3873-112-772-747} then ever before. The $\Lambda$CDM model solves this problem by assuming a much larger dark form of matter present in every galaxy and cluster of galaxies. However, over the decades no observation has confirmed the existence of this material either way \cite{Bertone:2004pz,Bi:2014hpa,Cirelli:2012tf}.\medskip

The unmodified Newtonian picture treats each galaxy as a collection of $N$ individual sources with a typical mass $M_{\odot}$, and then combines the respective Newtonian potentials. This generates a global outline of the Newtonian potential. Assuming a thin disk shaped galaxy with an exponential radial light distribution $\Sigma(R)=\Sigma_0 e^{-R/\beta_d}$, where $\beta_d$ is the disk scale radius and $\Sigma_0$ characterizes the central surface brightness, the resulting circular velocity for a test particle at a radial distance $R$ from the center of the disk is given by the Freeman formula \cite{Freeman1970,McGaugh:2014xfa}

\begin{equation}\label{eq.velGrdisk}
\begin{array}{l@{\hspace{0.3mm}}l@{\hspace{0.1mm}}l}
{v_{\scaleto{e}{3pt}_{{}_{\scaleto{TEGRd}{2pt}}}}}^2=&\dfrac{N M_{\odot} G R^2}{2 \beta_d^3}\left[I_0\left(\frac{R}{2 \beta_d}\right)K_0\left(\frac{R}{2 \beta_d}\right)\right.\\
&\;\;\;\;\;\;\;\;\;\;\;\;\;\;\;\;\;\;\;\left.-I_1\left(\frac{R}{2 \beta_d}\right)K_1\left(\frac{R}{2 \beta_d}\right)\right],\\\\
\end{array}
\end{equation}
where $G$ is Newton's constant of gravity, and $\text{I}_n$ and $\text{K}_n$ ($n=0,1$) are modified Bessel functions of the first and second kind respectively \cite{watson1995treatise}. \medskip

The central surface brightness distribution can not be expressed by the disk on its own. As such another component of the galaxy, the bulge component, is defined. These bulges have a higher stellar concentration than the stellar disk. The bulge mass distribution is approximately spherically symmetric, with stellar orbits being roughly circular. Thus the individual sources would follow the velocity relation \cite{Sofue:2008wt}
\begin{equation}
V_{e_{TEGR_b}}=\sqrt{\frac{GM_b(R)}{R}},
\end{equation}
where the mass, $M_b(R)$, is determined through the surface density calculation \cite{1987gady.book.....B}
\begin{equation}\label{de_Vaucouleurs_dentiy}
\rho(R)=\frac{1}{\pi}\displaystyle\int_{R}^{\infty} \frac{d\Sigma_{b}(x)}{dx} \frac{1}{\sqrt{x^2-R^2}} dx,
\end{equation}
and $\Sigma_b(R)$ is the de Vaucouleurs profile for the surface mass density
\begin{equation}\label{de_Vaucouleurs_profile}
\Sigma_{\text{b}}(R)=\Sigma_{\text{be}} \text{Exp}\left[-\kappa\left(\left(\frac{R}{R_b}\right)^{1/4}-1\right)\right],
\end{equation}
with $\kappa=7.6695$, $\Sigma_{\text{be}}=3.2\times10^{3}\, \text{M}_{\odot} \text{pc}^{-2}$, and $R_b=0.5\,\text{kpc}$. In the following work, we follow suit and treat these two observable regions separately. \medskip

Teleparallel gravity is one alternative to GR where the mechanism by which gravitation is communicated is torsional rather than curvature based. This theory was first proposed by Einstein himself in order to unify gravitation and electromagnetism \cite{einstein1923}. The equivalent reformulation remained dormant until the relatively recent resurgence in alternative and modified theories of gravity. Now, the resulting theory is equivalent at the level of equations and phenomenology. However, the resulting action can be generalized much like the $f(R)$ proposition. This is the origin of the distinction between the two theories in that the equivalence no longer holds. In fact, while the $f(R)$ field equations are fourth order, the $f(T)$ field equations are second order. \medskip

Galactic rotation curves were first tackled in $f(T)$ gravity in Ref.\cite{Jamil:2012ju}. While this study provides promising results there are some issues with the procedure and theory. Firstly, in the work, the separate regions of galaxies are not treated individually. Secondly, the mass profile is considered to be spherically symmetric. This may be true for the surrounding dark matter halo but not for the luminous matter segment of the galaxy. Finally, the choice of tetrad, $\udt{e}{a}{\mu}$, for the metric should be treated with somewhat different field equations. We expand on this in section II but the main point is that certain choices of tetrad require an associated quantity (spin connection) to be defined so that the theory continues to respect local Lorentz invariance, as explained in Ref.\cite{Krssak:2015oua}. However, the work does result in flat rotation disks in general. \medskip

The paper is divided as follows. In section II we give a brief overview of $f(T)$ gravity with an emphasis on the relevant solution to the field equations. In section III the mechanics of galactic rotation curves are worked through for the current setting. The resulting dynamics are then applied to Milky Way data for several instances of the general model under consideration in section IV. The best model is highlighted, and then a three-parameter fitting for the bulge and disk masses as well as the new coupling parameter is described and determined. In section V, we tackle the broader problem of galaxies with other morphologies. Finally, the results are discussed in section VI.

\section{II. $f(T)$ Gravity}
GR and its modifications are largely based on the metric tensor, $g_{\mu\nu}$, which solves the field equations and acts as the fundamental dynamical variable. The metric acts as a potential quantity with curvature being represented through the Levi-Civita connection (torsion-free), $\Gamma^{\lambda}_{\mu\nu}$. In teleparallelism this connection is replaced by the Weitzenb\"{o}ck connection, $\hat{\Gamma}^{\lambda}_{\mu\nu}$. This new connection is curvature-free and is based on two fundamental dynamical variables, namely the tetrads (or vierbein) and the spin connection. The tetrads, $\udt{e}{a}{\mu}$, are four orthonormal vectors that transform inertial and global frames in that they build the metric up from the Minkowski metric by means of an application of this transformation. They also observe the metricity condition. Physically, they represent the observer and can be related to the metric tensor by means of \cite{Saridakis2016}
\begin{equation}
g_{\mu\nu}=\eta_{ab}\udt{e}{a}{\mu}\udt{e}{b}{\nu}.
\end{equation}
where $\eta_{ab}=\text{diag}(1,-1,-1,-1)$. The tetrads obey the following inverse relations
\begin{equation}
\udt{e}{a}{\mu}\dut{e}{a}{\nu}=\delta^{\nu}_{\mu} \quad \udt{e}{a}{\mu}\dut{e}{b}{\mu}=\delta^{a}_{b}.
\end{equation}
These conditions are not enough to fully constrain the tetrad, and so there is an element of choice in forming the tetrad frames. \medskip

The other necessary ingredient to describe $f(T)$ gravity is the spin connection, $\udt{\omega}{b}{a\mu}$ \cite{aldrovandi2012teleparallel,Krssak:2015oua}. This is not a tensor and its particular form depends heavily on the system under consideration, that is, it accounts for the coordinate system such that the theory remains covariant. The choice of tetrad plays a deciding factor in whether the spin connection vanishes or not. This leads to a division in tetrads \cite{Tamanini:2012hg}, there are {\it pure} tetrads whose associated spin connection vanishes, and {\it impure} tetrads who spin connection gives a nonzero contribution. In the current case, we will consider pure tetrads and so will not consider the spin connection any further. \medskip

With the introduction of the spin connection, the tetrad remains the fundamental dynamical field on the manifold since every tetrad ansatz produces a well-defined associated spin connection. The question of the inheritability of solutions in GR to teleparallel gravity is then resolved since TEGR is equivalent to GR at the level of equations \cite{Krssak:2015oua}. As vacuum GR solutions are also solutions to $f(R)$ gravity, TEGR (or GR) solutions are also solutions of $f(T)$ gravity. \medskip

In GR, curvature is communicated between tangent spaces through the Levi-Civita connection. Teleparallel gravity rests on the Weitzenb\"{o}ck connection which takes the form of \cite{Hayashi:1979qx}
\begin{equation}
\hat{\Gamma}^{\lambda}_{\mu\nu}=\dut{e}{a}{\lambda}\partial_{\mu}\udt{e}{a}{\nu}.
\end{equation}
This naturally leads to the torsion tensor \cite{Saridakis2016}
\begin{equation}
\udt{T}{\lambda}{\mu\nu}=\hat{\Gamma}^{\lambda}_{\mu\nu}-\hat{\Gamma}^{\lambda}_{\nu\mu}.
\end{equation}
While the resulting TEGR theory is equivalent to GR at the level of equations, the ingredients leading up to this are not. The difference between the Weitzenb\"{o}ck and the Levi-Civita connections is represented by the contorsion tensor
\begin{equation}
\udt{K}{\mu\nu}{a}=\frac{1}{2}\left(\dut{T}{a}{\mu\nu}+\udt{T}{\nu\mu}{a}-\udt{T}{\mu\nu}{a}\right).
\end{equation} \medskip

\noindent Lastly, the superpotential tensor is introduced
\begin{equation}
\dut{S}{a}{\mu\nu}=\udt{K}{\mu\nu}{a}-\dut{e}{a}{\nu}\udt{T}{\alpha\mu}{\alpha}+\dut{e}{a}{\mu}\udt{T}{\alpha\nu}{\alpha}.
\end{equation}
This is defined purely for convenience in the resulting equations \cite{Saridakis2016} but plays an important role in the gravitational energy-momentum tensor of teleparallel gravity. These tensors can be contracted to form the torsion scalar, $T=\udt{T}{a}{\mu\nu}\dut{S}{a}{\mu\nu}$, which is the Lagrangian for TEGR. \medskip

The distinction between GR and TEGR can now be made clearer, i.e. the relationship between the Ricci scalar, $R$ and the torsion scalar, $T$, can be expressed explicitly. The difference between the two quantities obviously lies in a boundary term since they produce the same theory at the level of equations \cite{Saridakis2016}. The distinction in the difference can be quantified through \cite{Bahamonde:2015zma}
\begin{equation}\label{eq bound}
R(e)=-T+B,
\end{equation}
where $B=\frac{2}{e}\partial_{\mu}\left(e\udt{T}{\lambda}{\lambda\mu}\right)=2\nabla_{\mu}\udt{T}{\lambda}{\lambda\mu}$ is the boundary term. This relation represents the source of the disparity between the $f(R)$ and $f(T)$ generalizations since the boundary term no longer remains a total divergence term when the generalization is taken. Thus, GR and TEGR would be indistinguishable in terms of observations. However, while astrophysical observations cannot test distinctions between curvature and torsion, they can compare the predictions of the models available. Once enough tests have been compiled, a multi-test survey would be the best way to compare and contrast the individual models of the two theories. \medskip

Thus, taking the Lagrangian $-T+B$ will precisely reproduce the Ricci scalar. As with the generalization of the GR Lagrangian to the $f(R)$ class of theories \cite{Sotiriou:2008rp,Capozziello:2011et}, TEGR can also be generalized to $f(T)$. However, the relation between the resulting theories stops being equivalent since $f(R)\neq f(-T)+B$ ($f(R)=f(-T+B)$). In fact, the ensuing field equations are unique in that, out of the three possible quantities involved, namely $R$, $T$, and $B$, $f(T)$ is the only Lagrangian that produces second order field equations \cite{Bahamonde:2015zma,Bahamonde:2016cul}. \medskip

Therefore the action with arbitrary functional form of the torsion scalar, $f(T)$, is given by
\begin{equation}
S=\frac{1}{4\tilde{\kappa}}\displaystyle\int d^4 x e f(T),
\label{torsion_action}
\end{equation}
where $\tilde{\kappa}=4\pi G$ and $e=\text{det}\left(\udt{e}{a}{\mu}\right)$. Taking a variable with respect to the tetrad results in the field equations \cite{Saridakis2016}
\begin{align}
\dut{E}{a}{\mu}&\equiv  e^{-1} f_{T} \partial_{\nu}\left(e \dut{S}{a}{\mu\nu}\right)+f_{TT} \dut{S}{a}{\mu\nu} \partial_{\nu} T\nonumber\\
&-f_{T} \udt{T}{b}{\nu a}\dut{S}{b}{\nu\mu}+\frac{1}{4}f(T)\dut{e}{a}{\mu}=\tilde{\kappa} \dut{\Theta}{a}{\mu},
\end{align}
where $\dut{\Theta}{a}{\mu}\equiv \frac{1}{e}\frac{\delta \mathcal{L}_m}{\delta \udt{e}{a}{\mu}}$, $f_{T}$ and $f_{TT}$ denote the first and second derivatives of $f(T)$ with respect to $T$, and $\mathcal{L}_m$ is the matter Lagrangian. \medskip

In Ref.\cite{Matteo2015}, the power-law instance of the Lagrangian is investigated, i.e. $f(T)=T+\alpha T^n$ where $\alpha$ is a coupling constant and $|n|\neq1$ is any other real number. The following weak field solution is found
\begin{align}
ds^2 &= (1 + A(r))dt^2 - (1 + B(r))dr^{2} \nonumber\\
&- r^2d\theta^2 - r^2\sin^2(\theta)d\phi^2,
\label{eq_Matteos_metric}
\end{align}
where
\begin{equation}
A(r) = -\dfrac{2GM}{r}-\alpha\dfrac{r^{2-2n}}{2n-3}2^{3n-1},
\end{equation}
and
\begin{align}
B(r) =\dfrac{2GM}{r}+\alpha&\dfrac{r^{2-2n}}{2n-3}2^{3n-1}\\
&(-3n+1+2n^2).
\end{align}

For this case the torsion scalar is defined as follows

\begin{align}
T = &\frac{(-1-B(r))(3+B(r))}{r^2})\nonumber\\
&\hspace{0.5cm}\times[1+\frac{1+B(r)}{2} + r(1+A(r))A'(r)]
\end{align}

In the limit of vanishing $\alpha$, GR is again recovered. By determining the galactic rotation curve dynamics for this solution, we will investigate the effect of various $n$ values on the resulting behavior.

\section{III. Galactic Rotation Curves in $f(T)$}
In order to determine the rotational curve profile we consider a point particle with energy, $E$, and angular momentum, $L$, performing orbits about the galactic core. In the following we investigate this type of orbit with a focus on the effective potential. We then use this result to determine the velocity profile for the disk and the bulge separately.

\subsection{A. Effective Potential}
In order to obtain an effective potential we follow the procedure described in Wald \cite{wald1984general}. We start by determining the conservation relations for the energy and angular momentum from the background metric in Eq.(\ref{eq_Matteos_metric}). These are given by

\begin{equation}\label{eq_E}
E = -g_{\mu\nu}\zeta^{\mu}u^{\nu}= (1+A(r))\dfrac{dt}{d\tau},
\end{equation}
and
\begin{equation}\label{eq_L}
L = g_{\mu\nu}\Psi^{\mu}u^{\nu}= r^2\dfrac{d\varphi}{d\tau},
\end{equation}
where $\zeta^{\mu}=(\delta/\delta{t})^{\mu}$ and $\Psi^{\mu}=(\delta/\delta\phi)^{\mu}$ are the static and rotational killing vectors respectively. The background metric naturally leads to the radial differential relation below
\begin{equation}\label{eqn_of_motion}
\begin{array}{l@{\hspace{0.3mm}}l@{\hspace{0.1mm}}l}
1&=\dfrac{E^2}{(1 + A(r))} - (1+B(r))\left(\dfrac{dr}{d\tau}\right)^{2}-\dfrac{L^2}{r^2},\\\\
\end{array}
\end{equation}
where Eqs.(\ref{eq_E}--\ref{eq_L}) were used. \medskip

The effective potential, $V_e$, can be read off from Eq.(\ref{eqn_of_motion}) by comparison with Ref.\cite{wald1984general}
\begin{equation}
\left(\dfrac{dr}{d\tau}\right)^{2}=\dfrac{E^2}{2} - V_{e}.
\end{equation}
As a gravitational source, the system under consideration is not being treated as having any rotation, and so we can set $L=0$. Finally, if we assume roughly circular orbits, i.e. $\dfrac{dr}{d\tau} = 0$, and simplify the resulting expression, then we find an effective potential
\begin{align}\label{eq.effpot}
V_{e}&=\dfrac{(1 + A(r))}{2}\nonumber\\
&=\dfrac{1}{2}-\dfrac{GM}{r}-\alpha\dfrac{r^{2-2n}}{2n-3}2^{3n-2}.
\end{align}
Here we see that the effective potential includes the GR potential as well as an extra $f(T)$ contribution which can be divided as follows
\begin{equation}\label{eq.effGr}
V_{e_{{}_{TEGR}}}=\dfrac{1}{2}-\dfrac{GM}{r},
\end{equation}
and
\begin{equation}\label{eq.effalf}
\begin{array}{l@{\hspace{0.3mm}}l@{\hspace{0.1mm}}l}
V_{e_{{}_{\alpha}}}&=-\alpha\dfrac{r^{2-2n}}{2n-3}2^{3n-2}.
\end{array}
\end{equation}
Since the velocity contribution of the GR potential is known, we will now continue with the derivation for the $f(T)$ component. \medskip

With the effective potential in hand, it is now possible to obtain the velocity curve profile. To do this, we consider the centripetal and gravitational acceleration equations \cite{Toomre1963}
\begin{equation}\label{eq.centripetal acelleration}
a_c=\dfrac{v^2}{r},
\end{equation}
and
\begin{equation}\label{eq.Grav acelleration}
a_g=-\dfrac{dV_{eff}}{dr}.
\end{equation}

Assuming circular paths for the orbiting stars and dust, the effective velocity profile for the central mass turns out to be \cite{Freeman1970}

\begin{equation}\label{eq.pot1star}
{v_{eff}}^2=-r\dfrac{dV_{eff}}{dr},
\end{equation}
where the potential is inherently negative.

\subsection{B. Disk and Bulge}
Given the geometric diversity of the galactic disk and bulge regions, the two sectors are treated separately in the following calculations. In particular, the core contrast in the treatment is related to the difference in their mass density distributions which clearly affects the whole calculation due to the stark change in the effective potential.

\subsubsection{1. Disk}
\begin{figure}[h]
\includegraphics[width=8cm]{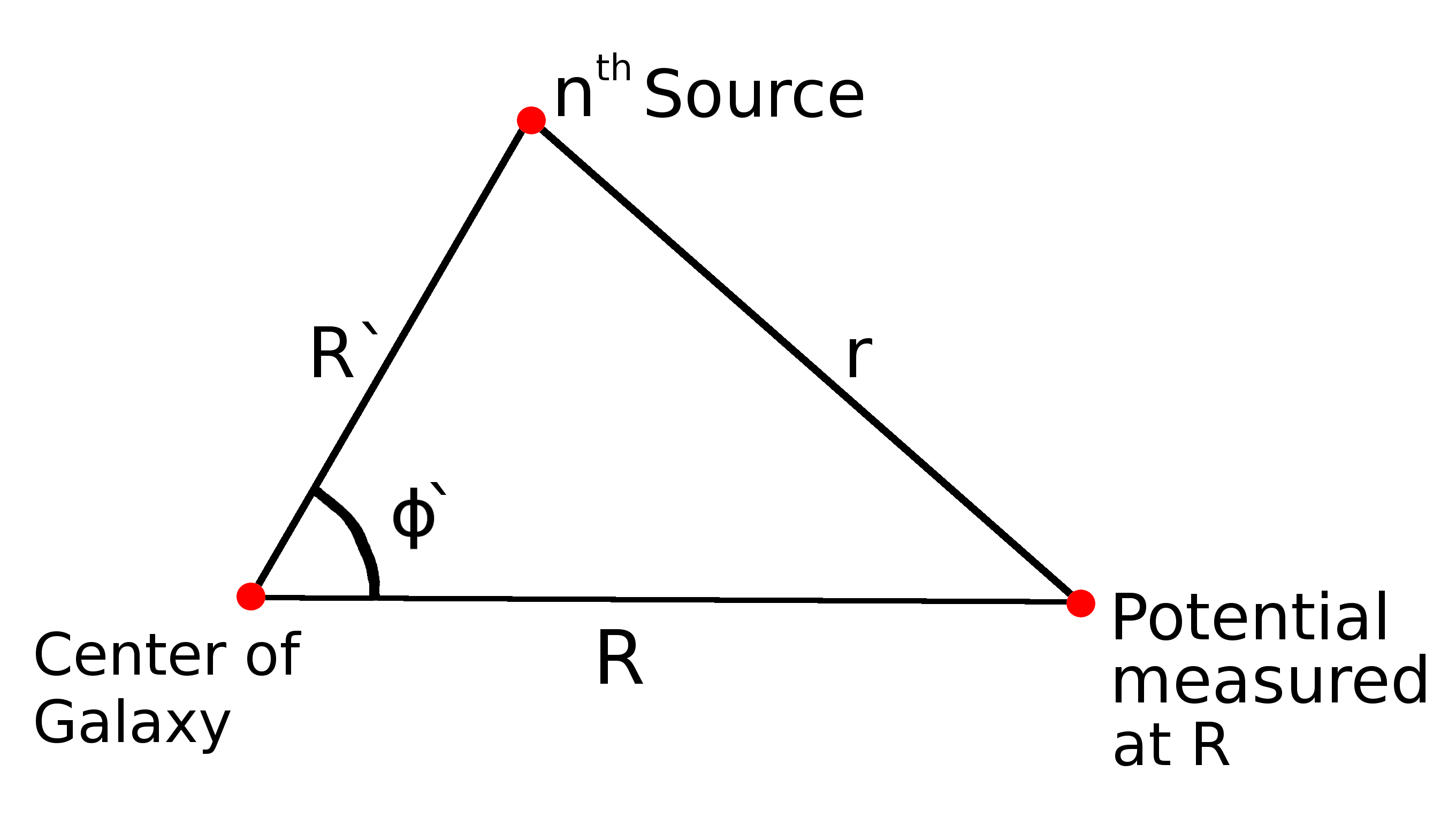}
\caption{For a position with radius $R$, the relative distances of an $n^{\text{th}}$ source are shown, where the radius of the source is denoted by $R'$, and the $r$ represents the distance between the source and the position where the potential is being measured. All distances are in cylindrical coordinates.}
\label{fig.cos}
\end{figure}

In order to determine the velocity curve profile of the disk component of galaxies we follow the method developed in \cite{1983MNRAS.203..735C,Milgrom:1983pn,Mannheim2006}. Consider a system of $N$ galactic bodies. The calculation of the velocity profile will necessarily involve the sum of the combined potential of the individual sources within the galaxy. To measure the potential for a particular position with radius $R$, all other source will be summed together. Consider the $n^{\text{th}}$ source with radius $R'$ from the galactic center; the distance between the position where the potential is being measured and the $n^{\text{th}}$ source will be denoted by $r$. This is depicted in Fig.(\ref{fig.cos}) where cylindrical coordinates are used.
 
With this picture in mind, we can proceed to express the unknown radius $r$ in terms of other radial terms as follows
\begin{align}\label{eq.cosrul}
r&=({R'}^2+{R}^2-2RR'\cos{(\phi-\phi')}\nonumber\\
&+(z-z')^{2})^\frac{1}{2},
\end{align}
where $z'$ is the height of the $n^{\text{th}}$ source, and $(\phi,z)$ represents an arbitrary reference position which we choose to take as the origin. \medskip

In order to obtain the combined potential of the whole disk we integrate over the whole range of source positions available
\begin{align}\label{eq.cylpot}
V_{e_{{}_{\alpha{d}}}}&=-\dfrac{{\alpha} 2^{3n-2}}{2n-3}{\displaystyle\int}\rho r^{2-2n}dV\nonumber\\
&=-\dfrac{{\alpha} 2^{3n-2}}{2n-3}{\displaystyle\int_0^{\infty}}dR'{\displaystyle\int_0^{2\pi}}d\phi'\nonumber\\
&\times{\displaystyle\int_{-\infty}^{\infty}}dz'R' \rho(R',z') r^{2-2n},
\end{align}
where $\rho$ is the mass density distribution given by \cite{Sofue2013}
\begin{equation}\label{diskdensit}
\rho(R',z')=\frac{\delta(z')M_0Ne^{-\frac{R'}{\beta_d}}}{\beta_d^22\pi},
\end{equation}
where $M_0$ represents one solar mass, $N$ is the total number of sources in the disk sector, and $\beta_d$ is the scale radius of the galactic disk. \medskip

It is at this point that the index $n$ cannot be left arbitrary in value, i.e. we must consider a set of values in order to proceed. The division is as follows: Integer values in the range $-\infty < n < 0$, $n=0$, all values in the ranges $0 < n < 1$ and $1 < n < \frac{3}{2}$. The core of the problem has to do with the expansion of the radial factor in the last integral of Eq.(\ref{eq.cylpot}). For the instances of integer values in the range $-\infty < n < 0$ and $n=0$, this results in the integral
\begin{align}\label{eq.dfinalGen}
V_{e_{{}_{\alpha{d}}}}&=-\dfrac{{\alpha} 2^{3n-2}}{2n-3}{\displaystyle\int_0^{\infty}}dR'{\displaystyle\int_0^{2\pi}}d\phi'{\displaystyle\int_{-\infty}^{\infty}}dz'R'\nonumber\\
&\times\dfrac{\delta(z')M_0Ne^{-\frac{R'}{\beta_d}}}{\beta_d^22\pi}\left[{R'}^2+{R}^2\right.\nonumber\\
&\left.-2RR'\cos{(\phi-\phi')}+(z-z')^{2}\right]^{1-n}.\nonumber\\
\end{align}

The calculations for the other ranges are presented in the appendices since they are more intricate. The resulting velocity curve profiles for the disk sector are given below for various $n$ values

$n=-4$ :
\begin{align}\label{eq.dveln4}
{v_{e_{{}_{\alpha{d}}}}}^2=&\dfrac{\alpha\;M_0\;N\;5\;R}{90112}\big(R^9+120R^7\beta_d^2+7200R^5\beta_d^4\nonumber\\
&+201600R^3\beta_d^6+1814400R\beta_d^8\big),
\end{align}

$n=-3$ :
\begin{align}\label{eq.dveln3}
{v_{e_{{}_{\alpha{d}}}}}^2=&\dfrac{\alpha\;M_0\;N\;R}{2304}\big(R^7+72R^5\beta_d^2+2160R^3\beta_d^4\nonumber\\
&+20160R\beta_d^6\big),
\end{align}

$n=-2$ :
\begin{equation}\label{eq.dveln2}
{v_{e_{{}_{\alpha{d}}}}}^2=\dfrac{\alpha\;3\;M_0\;N\;R}{896}\left(R^5+36R^3\beta_d^2+360R\beta_d^4\right),
\end{equation}

$n=0$ :
\begin{equation}\label{eq.dveln0}
{v_{e_{{}_{\alpha{d}}}}}^2=\dfrac{\alpha\;M_0\;N\;R^2}{6},
\end{equation}

$0 < n < 1$ :
\begin{align}\label{eq.dpot0_1}
{V_{e_{{}_{\alpha{d}}}}}=&-\dfrac{\alpha\;{2^{3n-2}}\;N\;M_0}{(2n-3)\;2\pi\;\beta_d^2}{\displaystyle\int_0^{\infty}}dR'R'e^{-\frac{R'}{{\beta}d}}\nonumber\\
&\Bigg(2\pi\;(R^{2}+R'^{2})^{-1-n}\Gamma(n)\Bigg\{(R^{2}+R'^{2})^{2}\nonumber\\
&\times{}_2F_1\left[\{\frac{n}{2},\frac{1+n}{2}\},\{1\},\frac{4R'^2R^2}{(R^{2}+R'^{2})^{2}}\right]\Bigg\}\nonumber\\
&-2nR'^2R^2{}_2F_1\Bigg[\{\frac{1+n}{2},\frac{2+n}{2}\},\{2\},\nonumber\\
&\frac{4R'^2R^2}{(R^{2}+R'^{2})^{2}}\Bigg]\Bigg).
\end{align}

This range of $n$ results in an integral that does not have a solution. Thus we solve it numerically in the next section. \medskip

$1 < n < \frac{3}{2}$ :
\begin{align}\label{eq.dvel1_3/2}
{v_{e_{{}_{\alpha{d}}}}}^2=&\dfrac{\alpha\;{8^{n-2}}\;N\;M_0R^2\beta_d^{-5-2n}\Gamma(4-2n)\Gamma(n-\frac{3}{2})}{(2n-3)\Gamma(n-1)}\nonumber\\
&\Bigg(4^n\sqrt{\pi}R^{3-2n}\beta_b^{2n}\Bigg[4(5-2n)\beta_b^2\nonumber\\
&\times\dfrac{{}_1F_2\left[\{\frac{3}{2}\},\{\frac{\pi}{2}-n,\frac{\pi}{2}-n\},\frac{R^2}{4\beta_d^2}\right]}{\Gamma(\frac{\pi}{2}-n)\Gamma(\frac{\pi}{2}-n)}\nonumber\\
&+3R^2\dfrac{{}_1F_2\left[\{\frac{5}{2}\},\{\frac{9}{2}-n,\frac{9}{2}-n\},\frac{R^2}{4\beta_d^2}\right]}{\Gamma(\frac{9}{2}-n)\Gamma(\frac{9}{2}-n)}\Bigg]\nonumber\\
&-128\beta_b^5\Gamma(n)\dfrac{{}_1F_2\left[\{n\},\{2,n-\frac{1}{2}\},\frac{R^2}{4\beta_d^2}\right]}{\Gamma(2)\Gamma(n-\frac{1}{2})}\Bigg).
\end{align}

\subsubsection{2. Bulge}
The bulge velocity profile calculation differs from the disk significantly in that it can be completed independently of the value of $n$. Following Refs.\cite{Sofue2008,Sofue2013}, we initially treat the bulge as a spherical mass which straightforwardly leads to
\begin{align}\label{eq.sing_pot}
{v_{e_{{}_{\alpha{b}}}}}^2&=R\;\frac{dV_{e_{{}_{\alpha{b}}}}}{dR}\nonumber\\
&=R\;\dfrac{d}{dR}\left(-\alpha\dfrac{R^{2-2n}}{2n-3}2^{3n-2}\right)M\nonumber\\
&=-\dfrac{{\alpha}\;2^{3n-2}(2-2n)}{(2n-3)}R^{2-2n}M.
\end{align}

For this region of the galaxy, the spherical mass distribution can be described through the de Vaucouleurs profile shown in Eq.(\ref{de_Vaucouleurs_profile}) which directly leads to the modified velocity profile
\begin{equation}\label{eq.bvel_preM}
{v_{e_{{}_{\alpha{b}}}}}^2=\scaleto{-\dfrac{{\alpha}\;2^{3n-2}(2-2n)}{(2n-3)}R^{2-2n}}{22pt}M(r),
\end{equation}
where the mass is calculated through the density distribution in Eq.(\ref{de_Vaucouleurs_dentiy}) and 
\begin{equation}
M(r)=\int{\rho(r)dV},
\end{equation}
which can turned into an integration over the individual spherical shells, giving
\begin{equation}
M(r)=4\pi\int_0^R{\rho(r){r^2}dr}.
\end{equation}

The contribution from the modified $f(T)$ gravity terms to the velocity profile then turns out to be 
\begin{align}\label{eq.bgen_vel}
{v_{e_{{}_{\alpha{b}}}}}^2&=\scaleto{\dfrac{{\alpha}\;2^{3n-2}(2-2n)}{(2n-3)}R^{2-2n}}{21pt}{\displaystyle\int_0^R}\nonumber\\
&{\times}\displaystyle{\int_r^{\infty}}\dfrac{d\Sigma_b(x)}{dx}\dfrac{r^2}{\sqrt{x^2-r^2}}dxdr\nonumber\\
&=\dfrac{{\alpha}\;2^{3n-2}(2-2n)}{(2n-3)}R^{2-2n}\nonumber\\
&\times\left(\frac{8388608\;{{\left(\frac{1}{\beta_b}\right)}^\frac{1}{4}}\kappa\;\Sigma_{be}\;{e^\kappa}\;Mg(R)}{\pi^3\;{\left(\frac{\kappa^8}{\beta_b^2}\right)}^{\frac{9}{8}}}\right),\nonumber\\
\end{align}
where $Mg(R)$ is the Meijer G--function defined through \cite{andrews1985special}
\begin{align}\label{eq.MeijG}
M&g(R) \equiv \nonumber\\
&G^{8,1}_{1,9}\Bigg(\frac{R^2\kappa^8}{16777216\beta_b^2}\:\:\Bigg|\:\:\begin{matrix}
1\\
\frac{9}{8},\frac{5}{4},\frac{11}{8},\frac{3}{2},\frac{3}{2},\frac{13}{8},\frac{7}{4},\frac{15}{8},0
\end{matrix}\Bigg).\nonumber\\
\end{align}

For the TEGR case, using the potential in Eq.(\ref{eq.effGr}) in the calculation in Eq.(\ref{eq.sing_pot}) leads directly to the velocity profile
\begin{align}\label{eq.velGrbulge}
{v_{e_{{}_{TEGRb}}}}^2=\dfrac{8388608\kappa G\;{{\left(\frac{1}{\beta_b}\right)}^\frac{1}{4}}\;\Sigma_{be}\;{e^\kappa}\;Mg(R)}{\pi^3\;R\;{\left(\frac{\kappa^8}{\beta_b^2}\right)}^{\frac{9}{8}}},
\end{align}
where $\beta_b$ is the bulge scalar radius \cite{Sofue2008}.

Now that both velocity profile sectors are derived, the full velocity profile can be derived for the $f(T)=T+\alpha{T^n}$ Lagrangian. This results in the velocity profile
\begin{equation}\label{eq.all_vect_vel}
v=\sqrt{{v_{e_{{}_{TEGRd}}}}^2+{v_{e_{{}_{TEGRb}}}}^2+{v_{e_{{}_{\alpha{b}}}}}^2+{v_{e_{{}_{\alpha{b}}}}}^2},
\end{equation}
where both TEGR and modified $f(T)$ gravity components are added for the disk and bulge segments of the galaxy.

\section{IV. The Milky Way Galaxy}
In the first part of this section we compare results for the velocity profile with Milky Way data using the combined velocity equations in Eq.(\ref{eq.all_vect_vel}) to determine the best range of values of $n$ in the Lagrangian, $f(T)=T+\alpha T^n$. At this stage, we simply fit for the constant $\alpha$. In the second part of the section the determined best range of $n$ will be used to fit for the surface mass density of the bulge $\Sigma_{be}$, the mass of the disk, $M$ and the coupling constant $\alpha$ to further test the velocity profile against real world data. \medskip

\subsection{A. Determination of best range of index $n$}
The data set being utilized is collected from two sources, namely Refs.\cite{Sofue2008,Data2}. A representative sample of this data is shown in Table \ref{Tab.Data}; this spans the breath of the region under consideration. Here, $R$ is the radial galactic distance shown in Eq.(\ref{eq.cosrul}), $v$ is the rotational velocity of the sources at that radial distance, and $e_u$ and $e_l$ are the upper and the lower error bars of the velocity values respectively. \medskip

In Table \ref{Tab.Data2} we present the necessary values of the constants to calculate the rotation curve profile contribution of the Milky Way bulge and disks for the various ranges of $n$ respectively \cite{Sofue2008}.

\onecolumngrid
\newpage
\begin{figure}[H]
\hspace{-1.3cm}\begin{minipage}{\textwidth}
\setlength{\tabcolsep}{-9.6pt} 
\begin{tabular}{ccc}
\includegraphics[width=70mm]{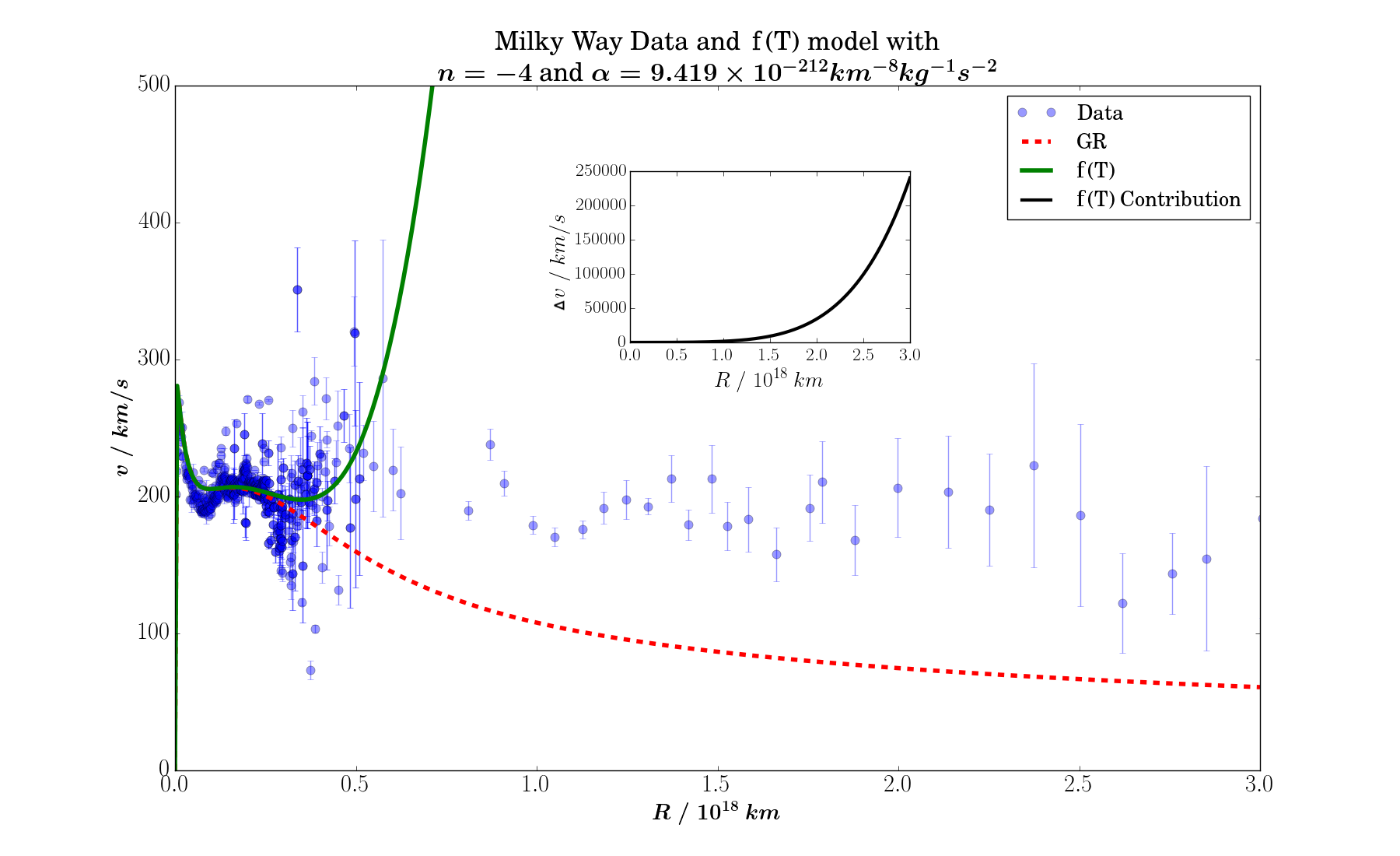}&
\includegraphics[width=70mm]{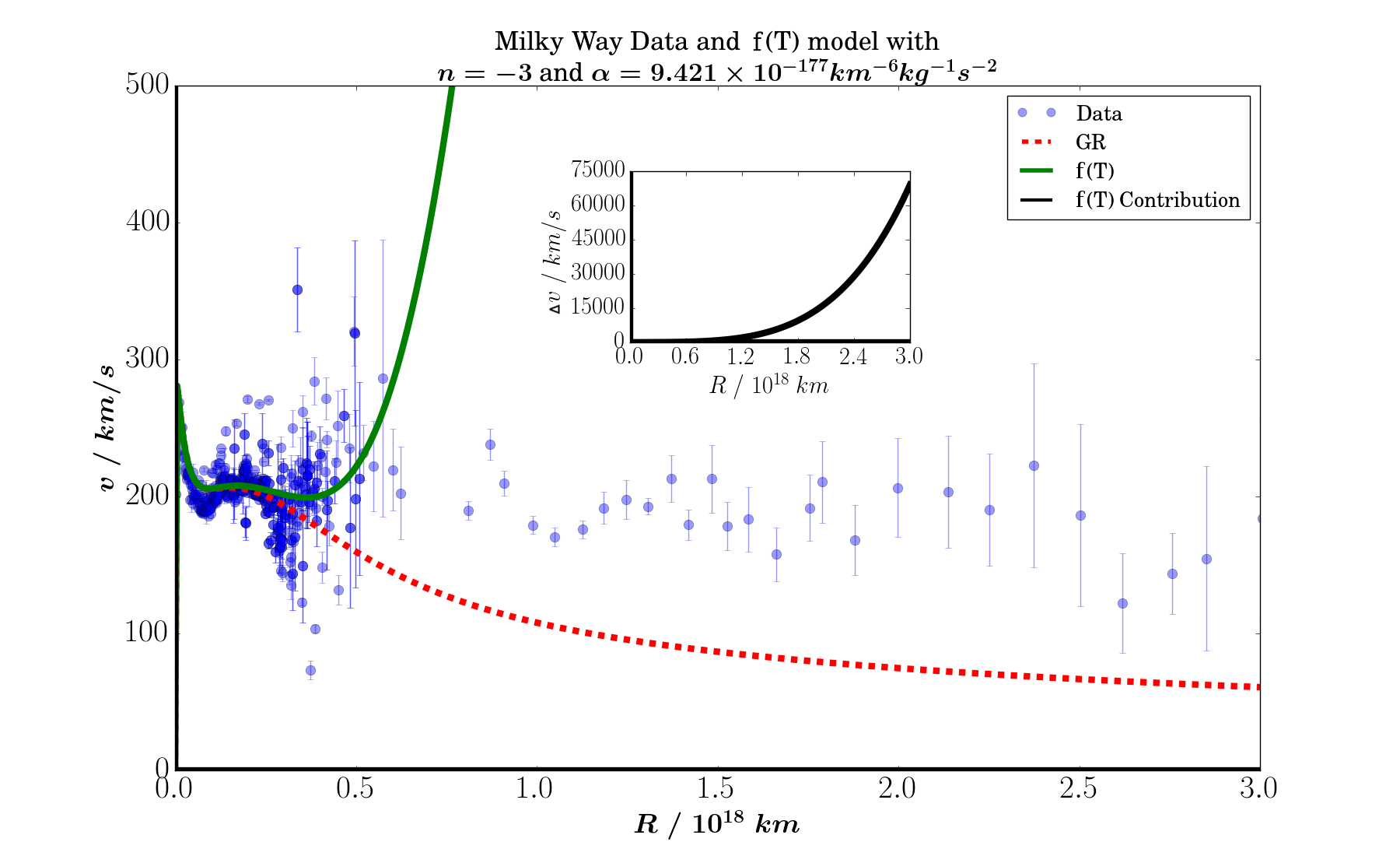}&
\includegraphics[width=70mm]{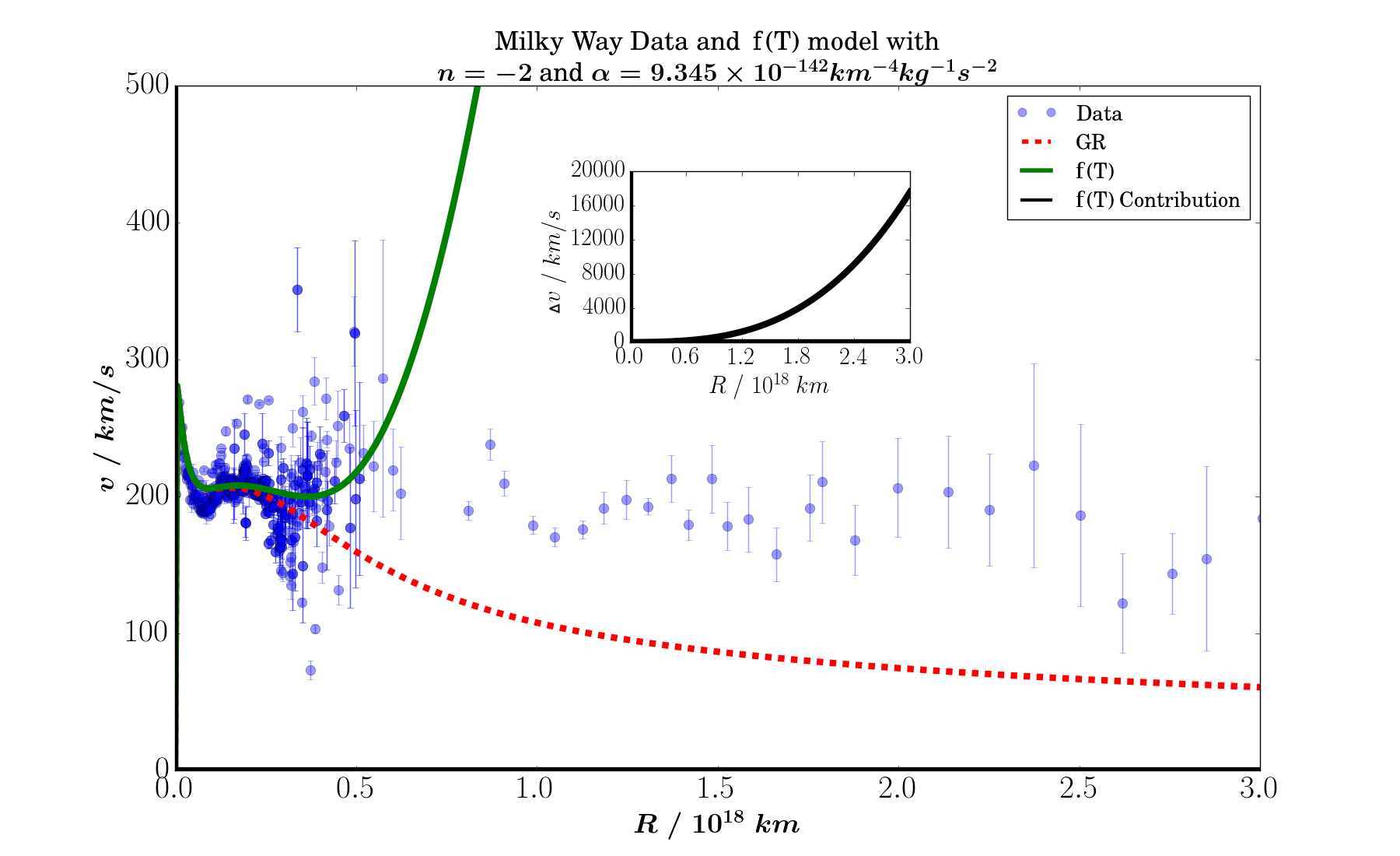}\\   
\includegraphics[width=70mm]{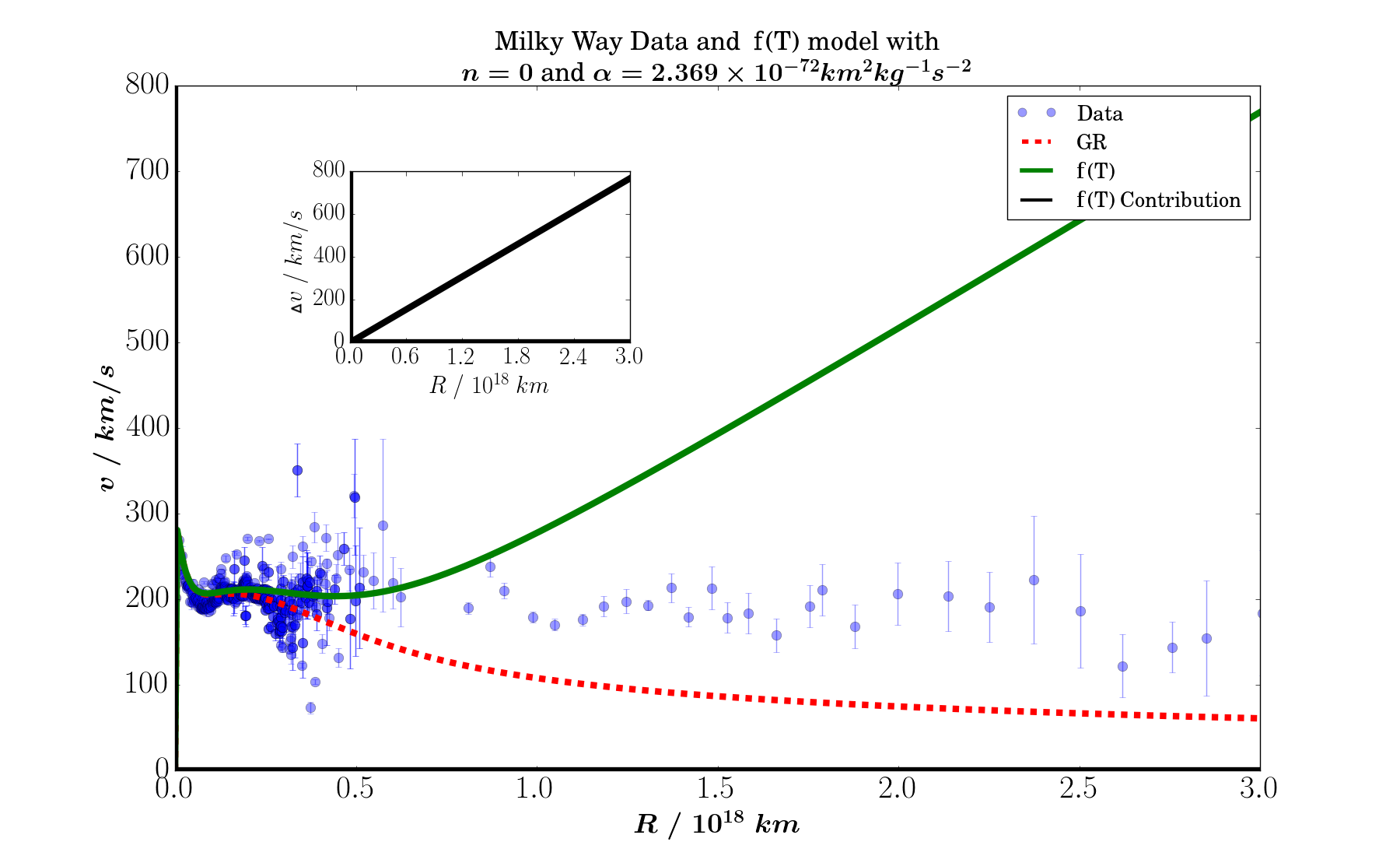}&
\includegraphics[width=70mm]{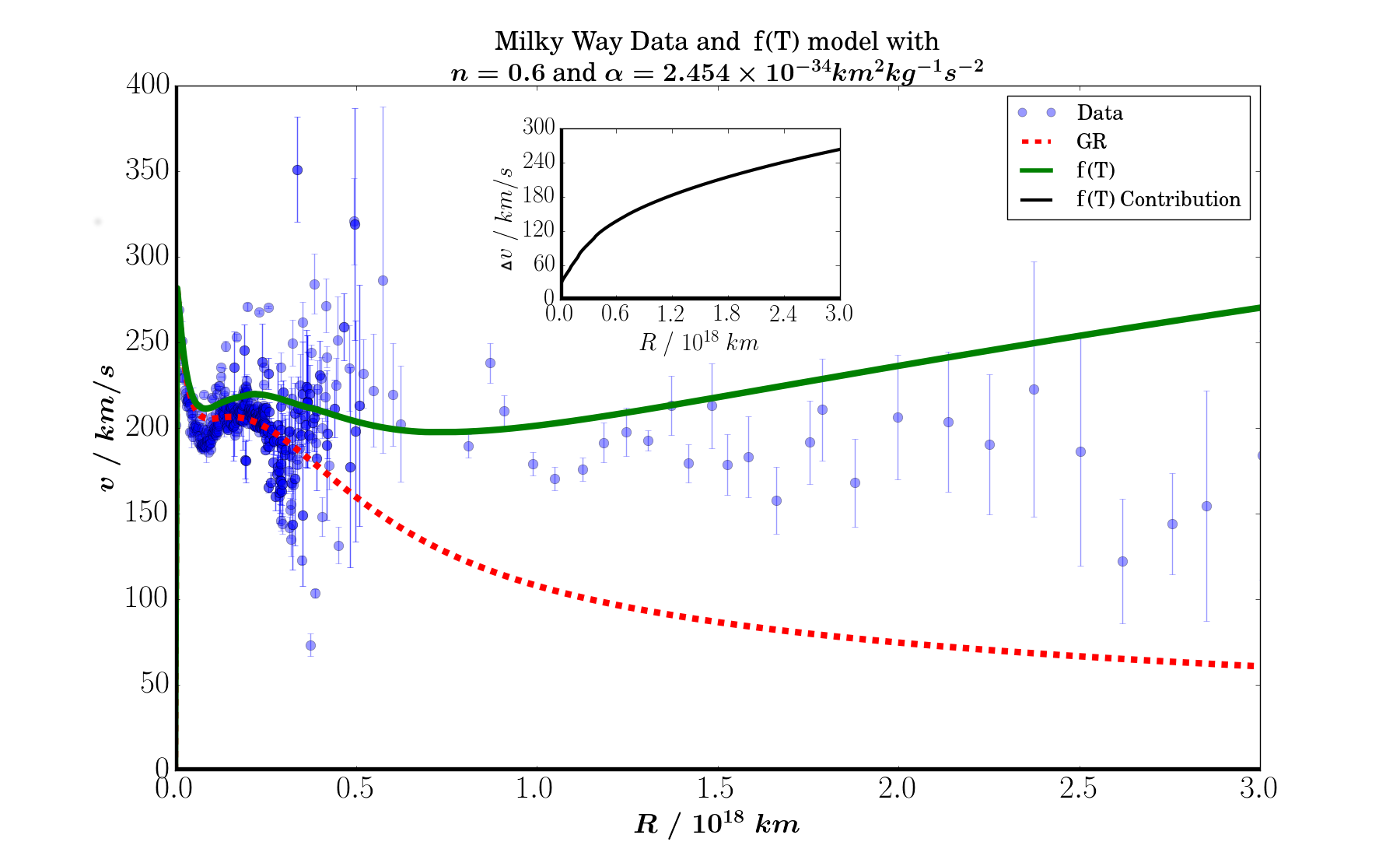}&
\includegraphics[width=70mm]{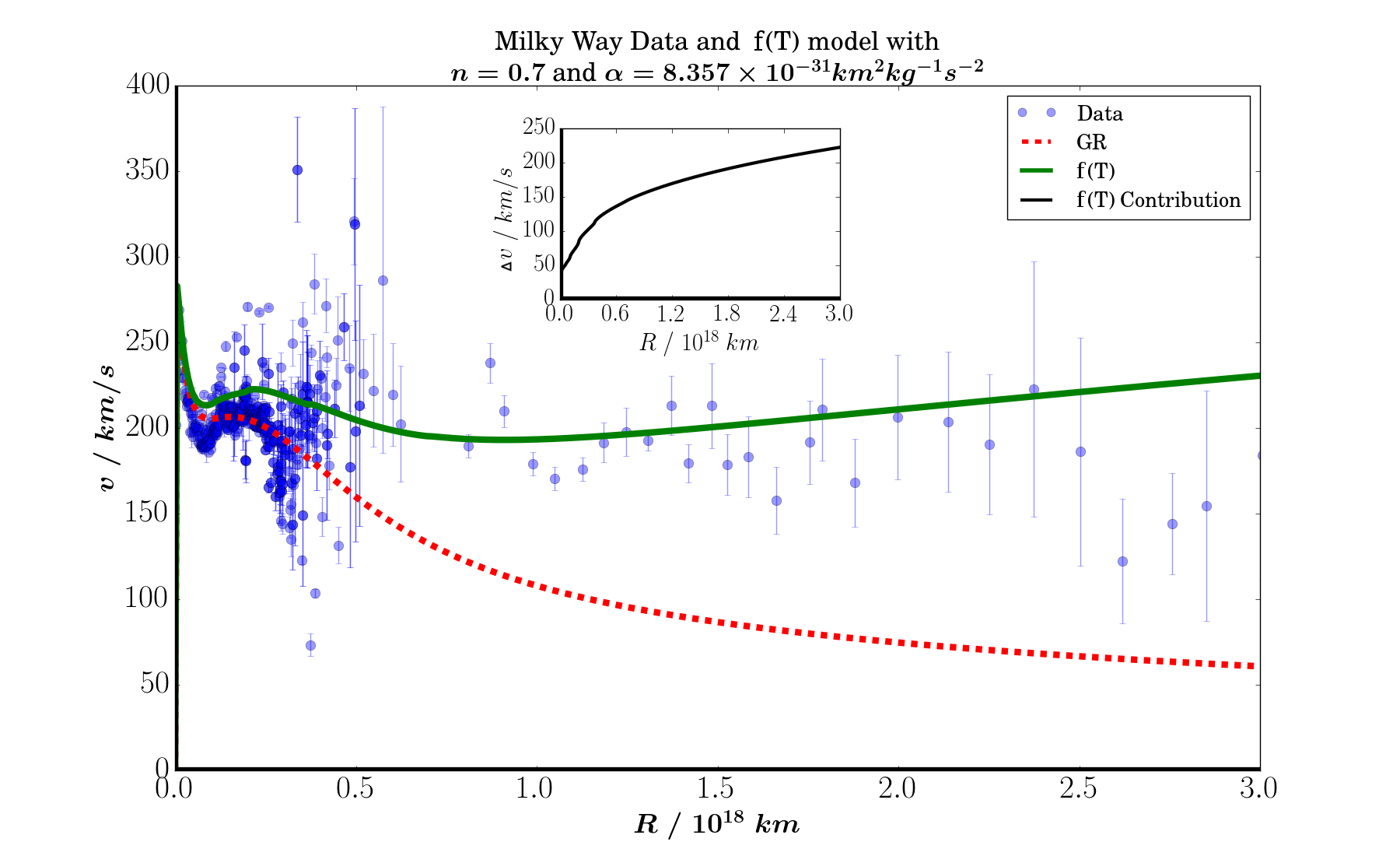}\\
\includegraphics[width=70mm]{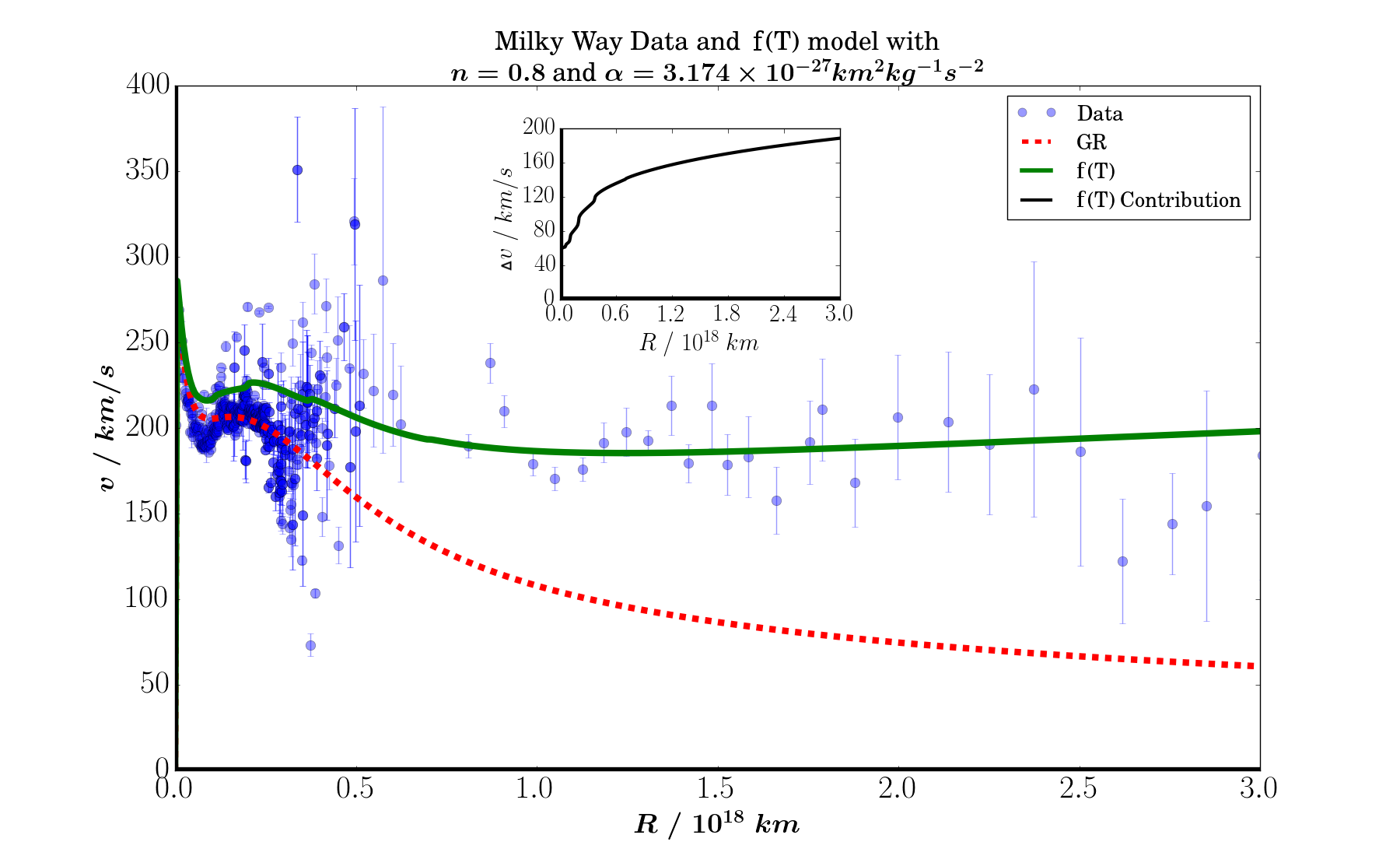}&
\includegraphics[width=70mm]{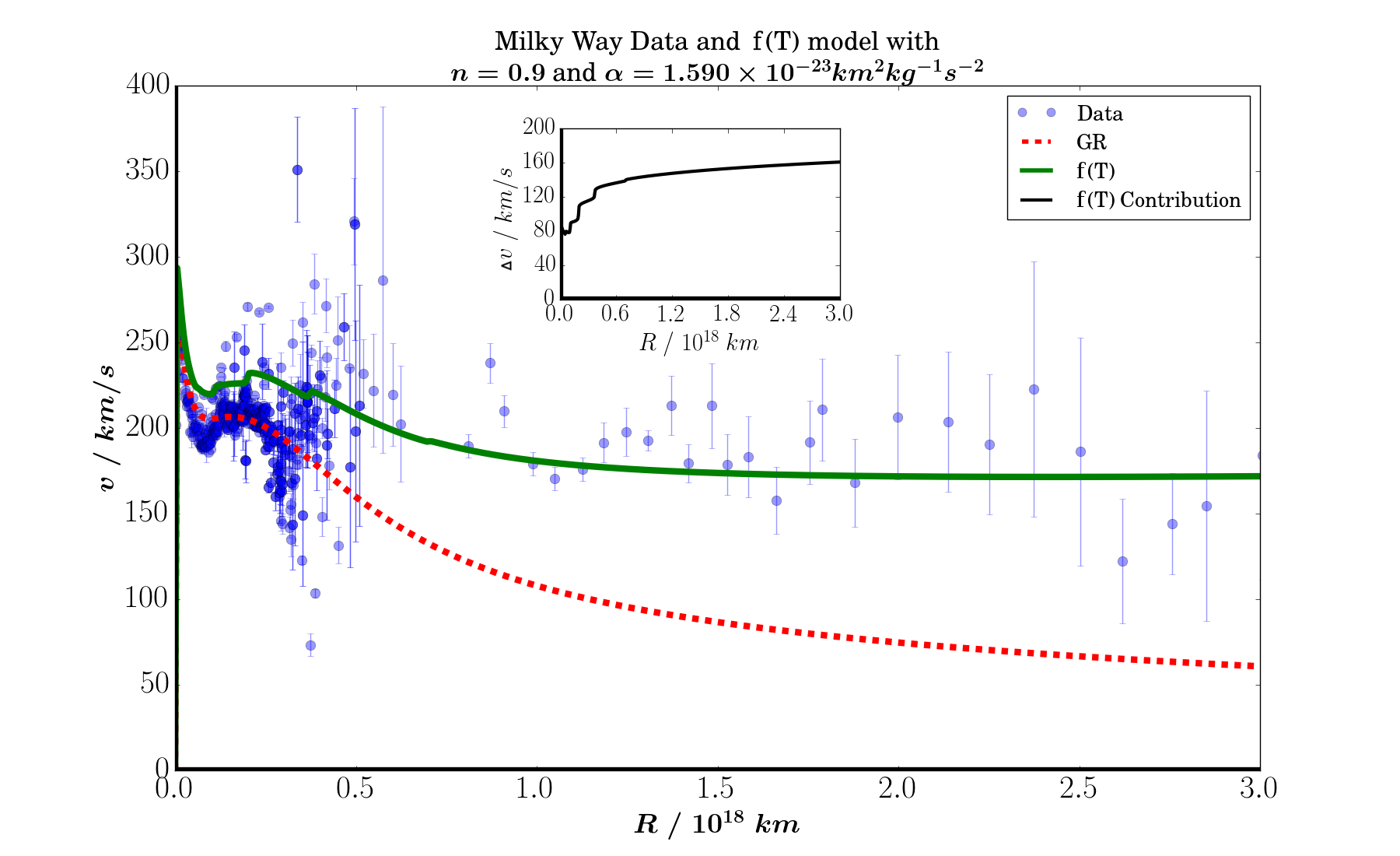}&
\includegraphics[width=70mm]{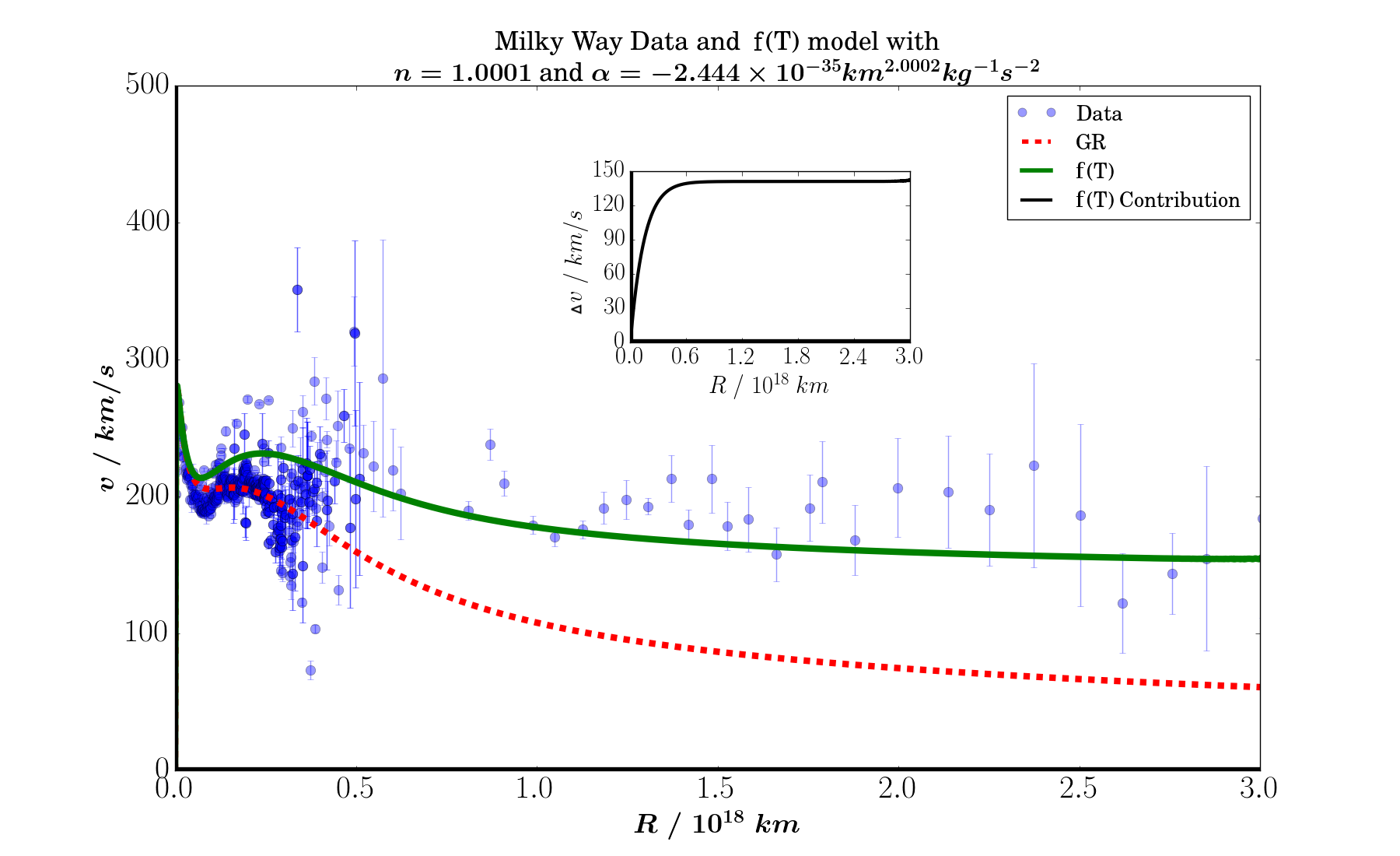}\\
\includegraphics[width=70mm]{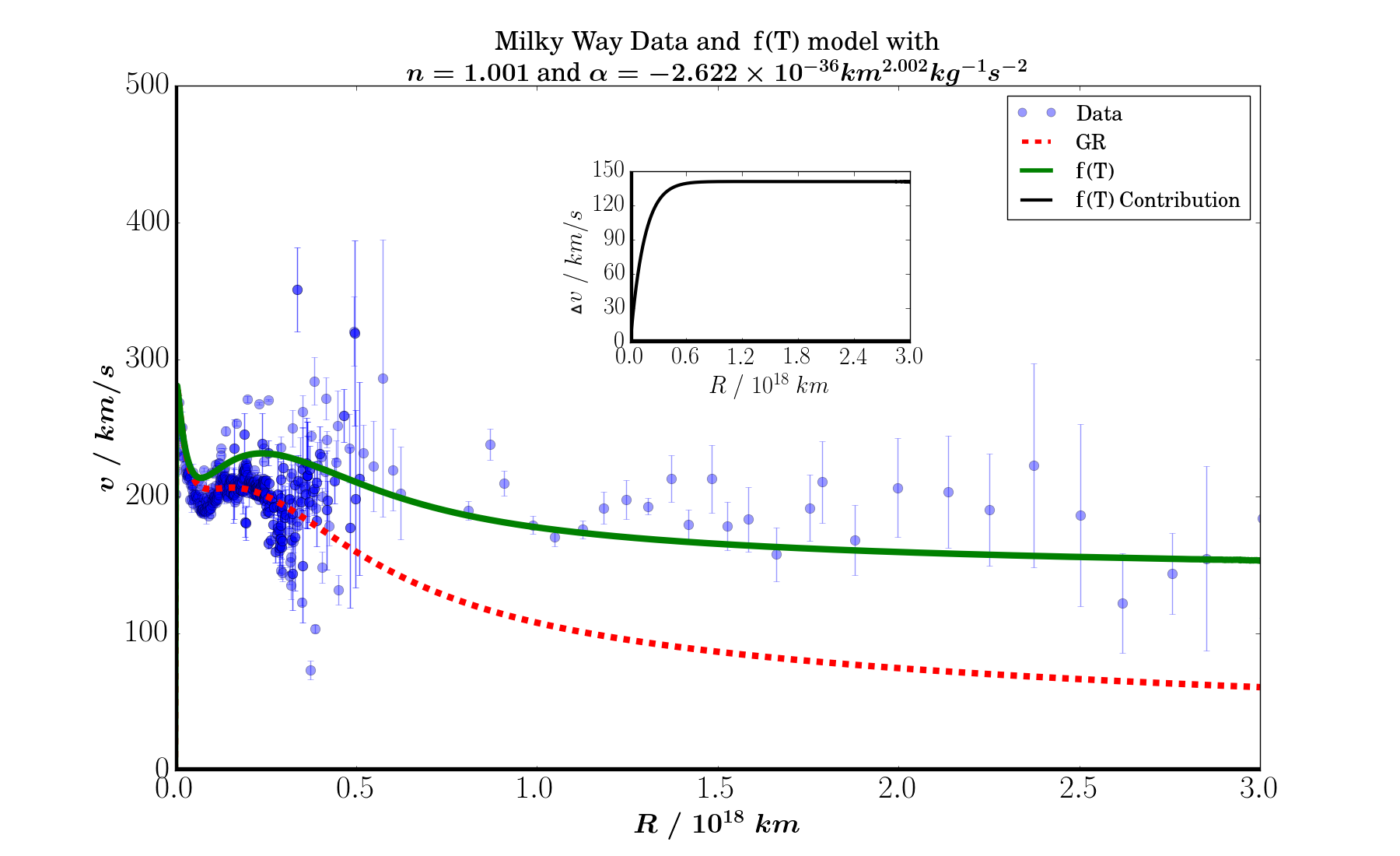}&
\includegraphics[width=70mm]{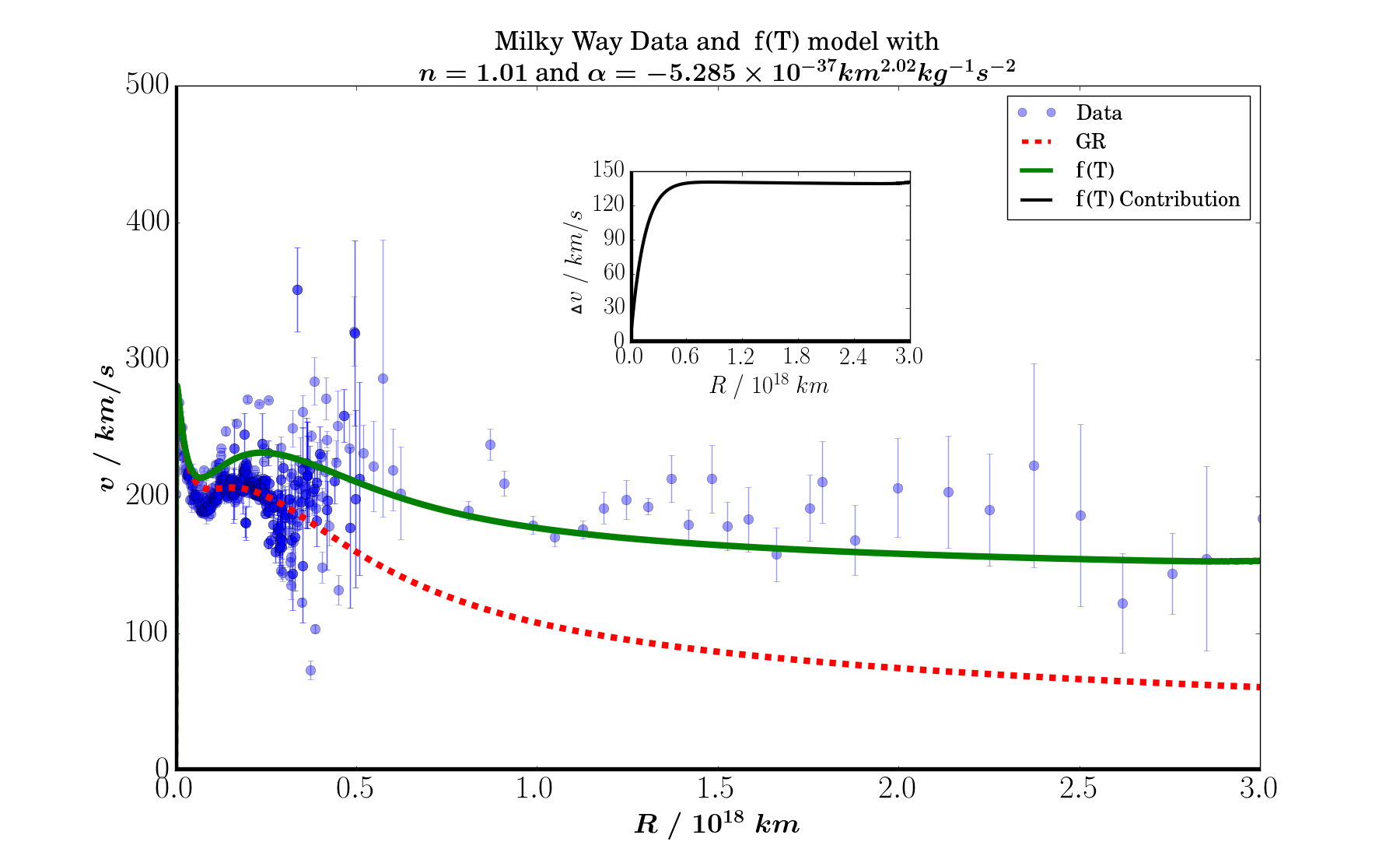}&
\includegraphics[width=70mm]{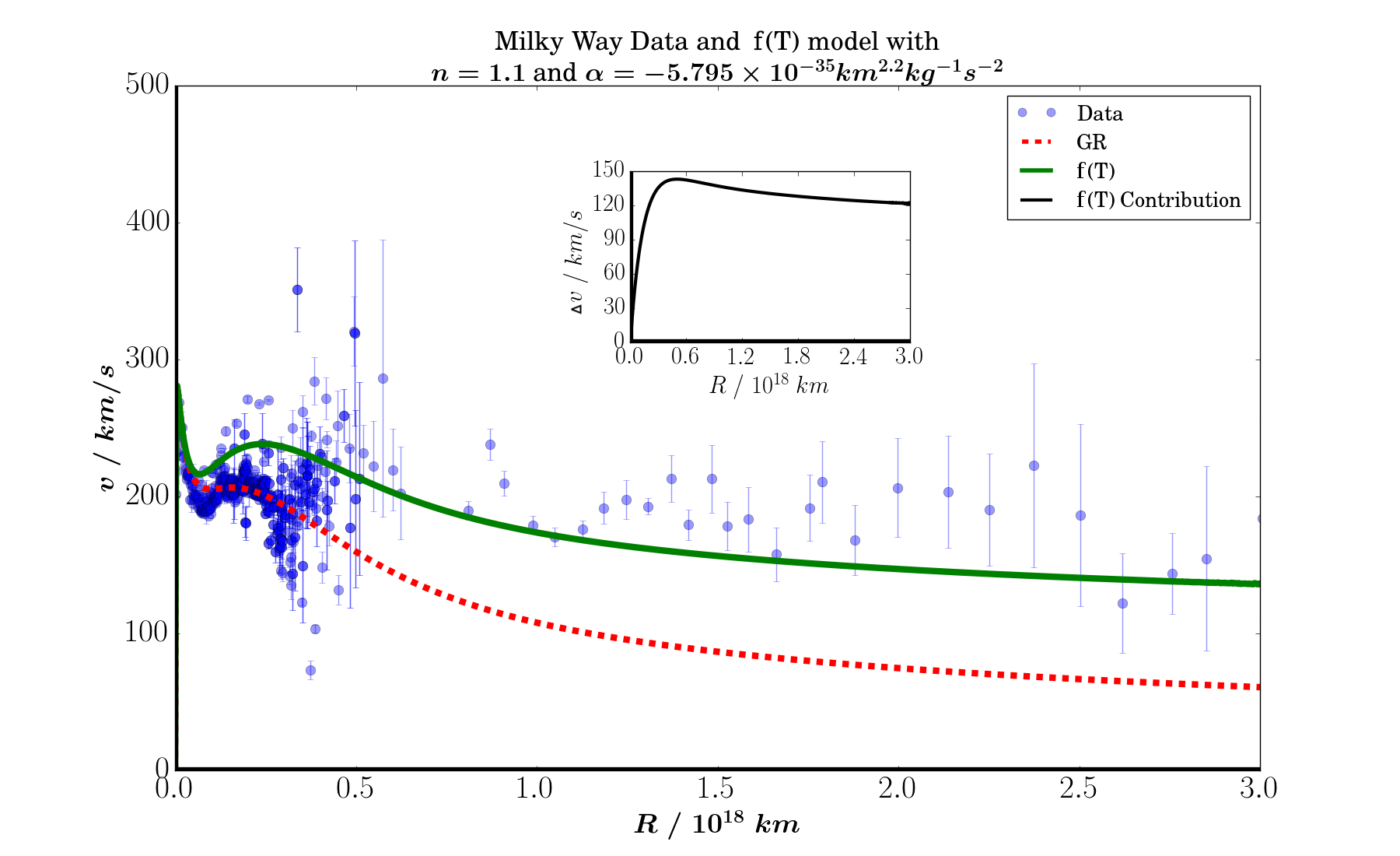}\\
\includegraphics[width=70mm]{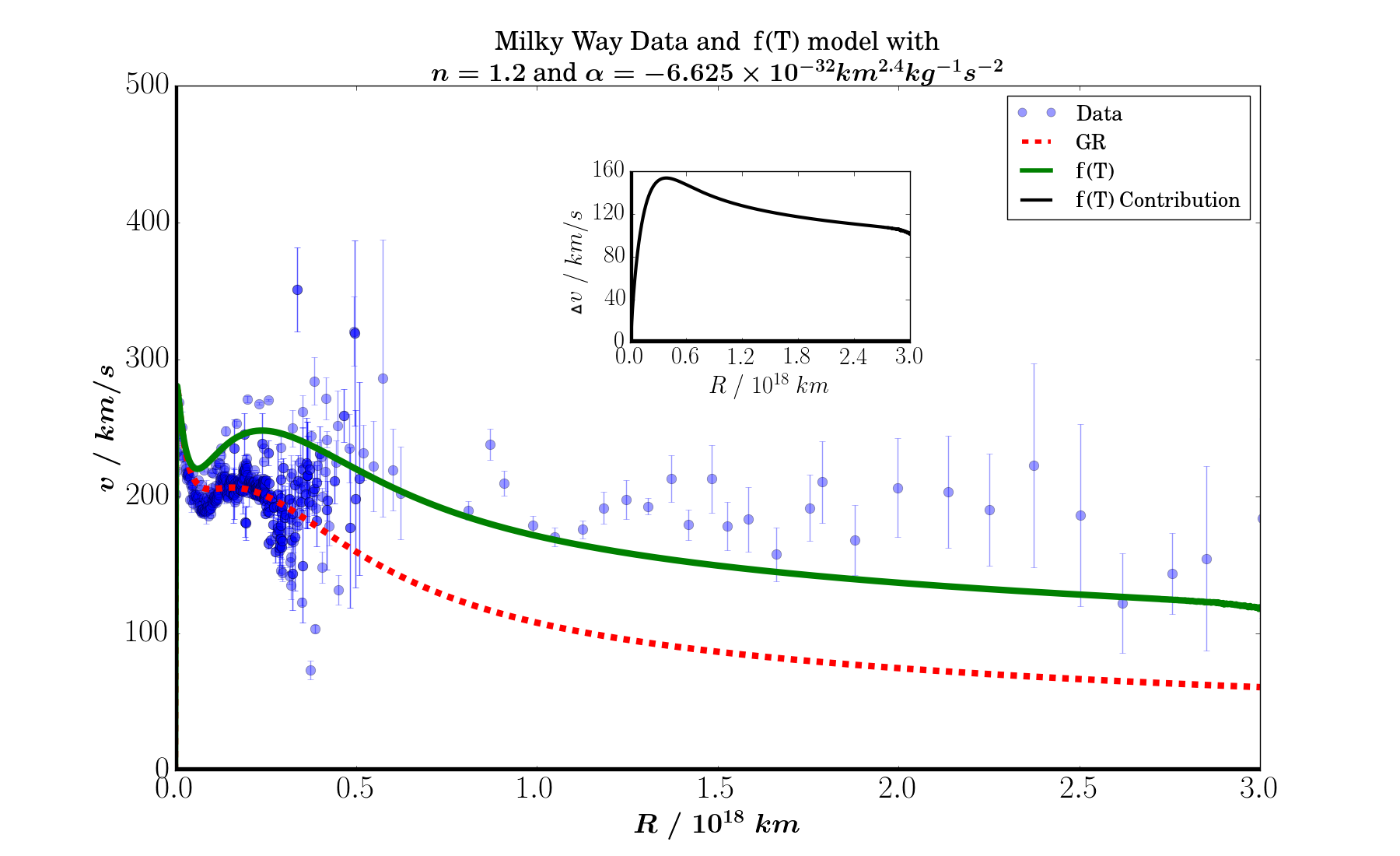}&
\includegraphics[width=70mm]{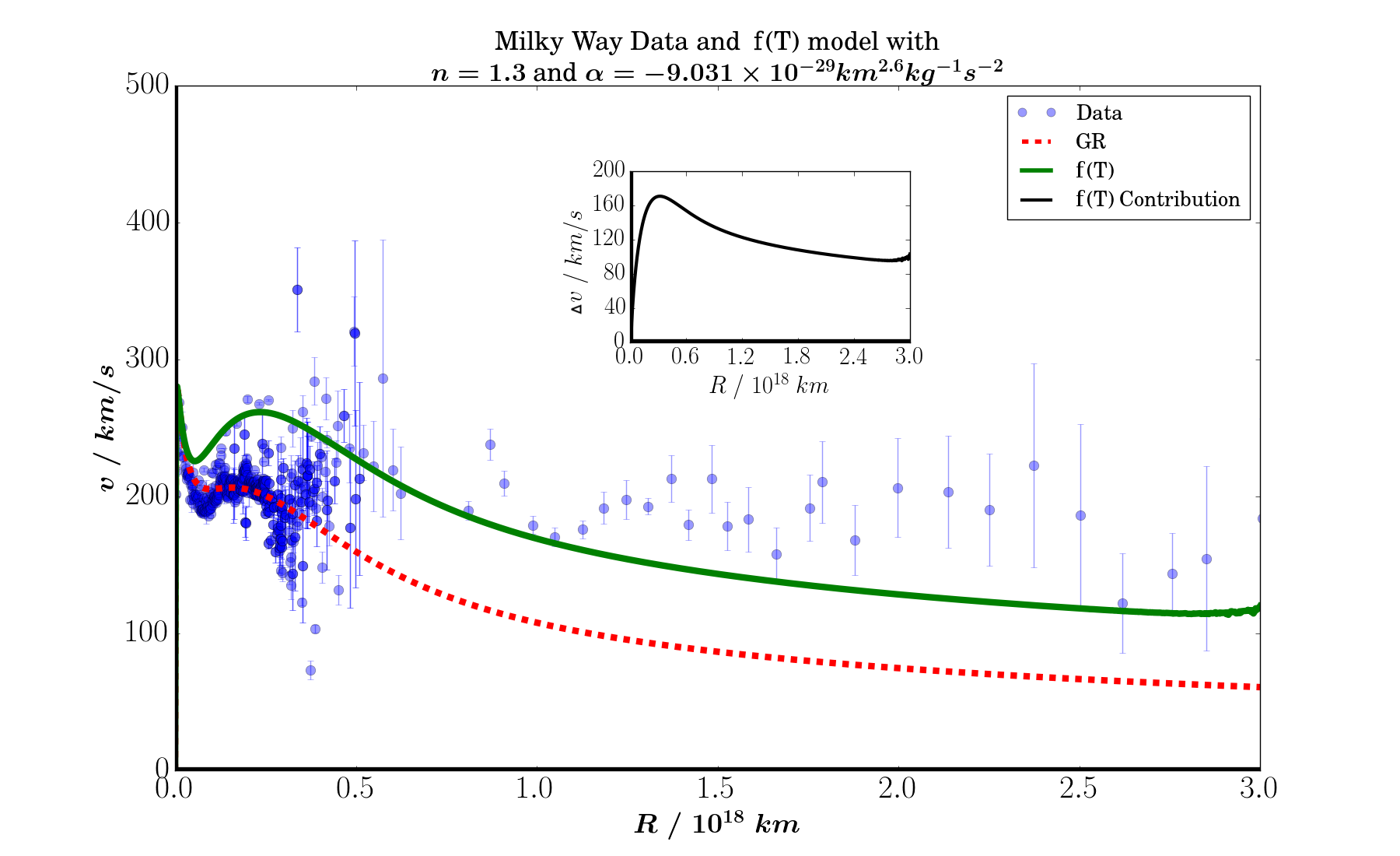}&
\includegraphics[width=70mm]{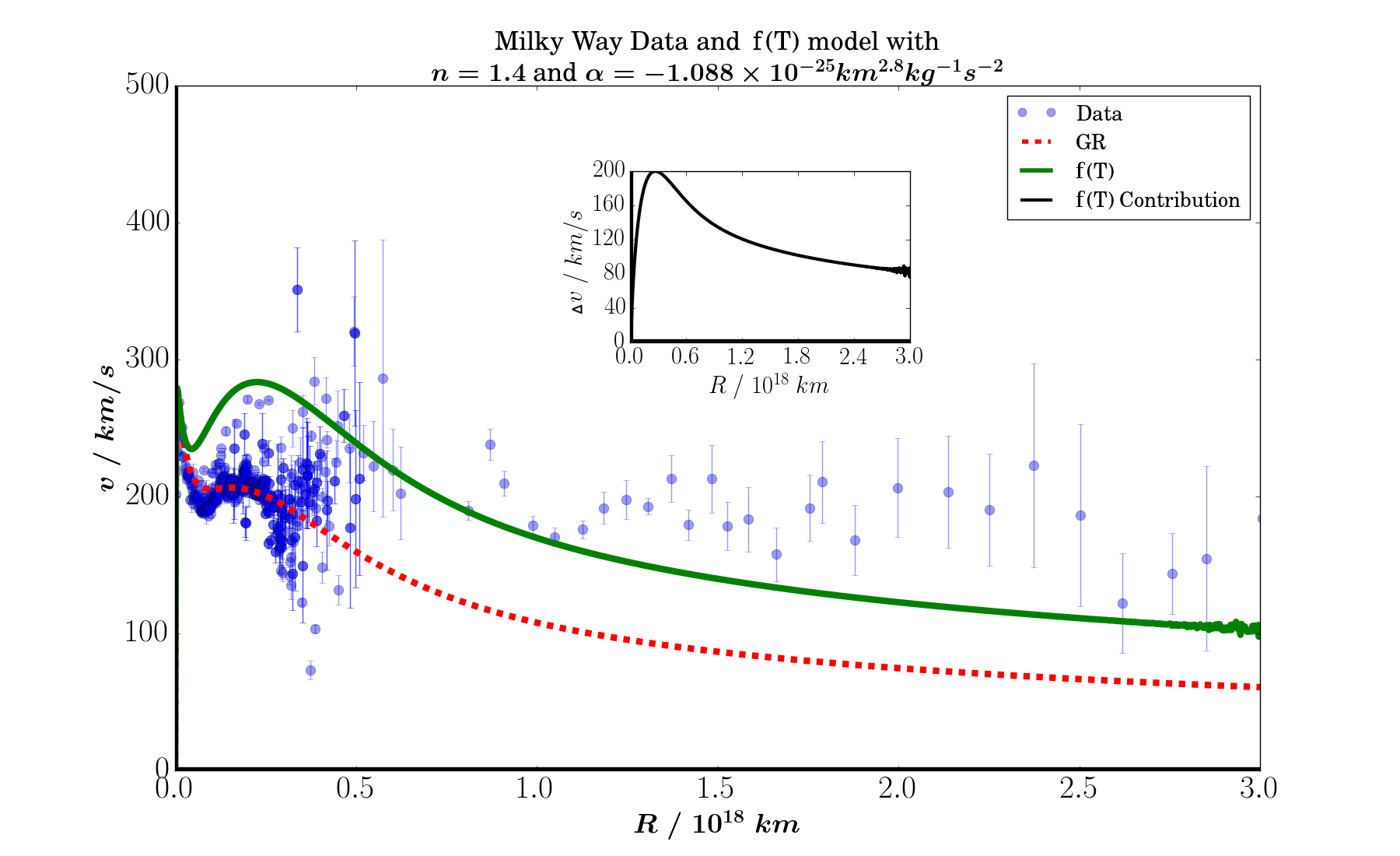}\\
\end{tabular}
\end{minipage}
\caption{$f(T)$ gravity rotational velocities with $n$ values from $-4.0$ to $1.4$. The coupling constant for each graph was obtained through fitting as governed by Eq.(\ref{eq fitting}). The data points are rotational velocity values obtained from a combination of two data sets, namely Refs.\cite{Sofue2008,Data2}. The red dashed curve represents the general relativistic rotation curve  while the full green curve represents the fitted $f(T)$ velocity curve with the resulting $\alpha$ value being indicated in every case. In all instances, the difference between the GR and the fitted $f(T)$ value is shown.}
\label{fig.grid}
\end{figure}

\begin{table*}[t]
\centering
\caption{Sample of the galactic rotation curve data. Column (1): radial distance from the galactic center. Column (2): velocity at that radial position. Column (3): upper error bar value. Column (4) lower error bar value. The columns then iterate for further data points}
\vspace{0.3cm}
\small
\setlength{\tabcolsep}{0.15cm} 
\hspace*{-0.2cm}\begin{tabular}{cccccccccccc}%
\hline
\multicolumn{1}{c}{$R$} & \multicolumn{1}{c}{${v}$} & \multicolumn{1}{c}{$e_u$} & \multicolumn{1}{c}{$e_l$} & \multicolumn{1}{c}{$R$} & \multicolumn{1}{c}{${v}$} & \multicolumn{1}{c}{$e_u$} & \multicolumn{1}{c}{$e_l$} & \multicolumn{1}{c}{$R$} & \multicolumn{1}{c}{${v}$} & \multicolumn{1}{c}{$e_u$} & \multicolumn{1}{c}{$e_l$}\\
\multicolumn{1}{c}{($10^{17}$km)} & \multicolumn{1}{c}{(km/s)} & \multicolumn{1}{c}{(km/s)} & \multicolumn{1}{c}{(km/s)} & \multicolumn{1}{c}{($10^{17}$km)} & \multicolumn{1}{c}{(km/s)} & \multicolumn{1}{c}{(km/s)} & \multicolumn{1}{c}{(km/s)} & \multicolumn{1}{c}{($10^{17}$km)} & \multicolumn{1}{c}{(km/s)} & \multicolumn{1}{c}{(km/s)} & \multicolumn{1}{c}{(km/s)}\\\hline
0.03&201.76&2.50&2.50&1.75&212.03&2.00&2.00&2.60&192.56&3.46&3.46\\
0.13&251.56&2.50&2.50&1.82&204.58&9.04&9.04&2.70&199.12&4.32&4.32\\
0.30&217.03&4.00&4.00&1.88&204.04&16.83&16.83&2.81&193.00&6.50&6.50\\
0.42&208.59&2.00&2.00&1.92&245.23&15.30&15.30&2.88&162.47&18.10&18.10\\
0.56&201.98&3.00&3.00&1.95&180.74&12.41&12.41&2.92&169.07&9.09&9.09\\
0.68&204.63&0.00&0.00&2.00&216.10&0.00&0.00&2.94&190.70&13.82&13.82\\
0.76&191.10&0.00&0.00&2.05&209.41&2.00&2.00&2.98&220.83&6.95&6.95\\
0.84&188.88&1.50&1.50&2.10&214.97&2.00&2.00&3.16&142.04&6.03&6.02\\
0.91&193.56&1.50&1.50&2.16&205.62&0.00&0.00&3.23&167.94&4.06&4.05\\
0.99&187.70&1.50&1.50&2.21&208.82&2.00&2.00&3.32&170.27&38.81&38.81\\
1.05&195.48&1.50&1.50&2.24&206.46&0.00&0.00&3.46&205.63&9.82&9.81\\
1.12&197.20&0.00&0.00&2.28&205.94&3.72&3.72&3.58&207.49&7.26&7.25\\
1.20&202.86&0.00&0.00&2.32&202.04&0.00&0.00&3.65&215.11&29.35&29.36\\
1.27&235.31&0.00&0.00&2.35&201.71&0.00&0.00&3.81&203.36&10.18&10.18\\
1.34&212.73&0.00&0.00&2.38&207.99&0.00&0.00&3.93&222.50&17.00&17.00\\
1.39&212.71&1.50&1.50&2.42&238.75&22.00&22.00&4.20&241.58&5.92&5.92\\
1.48&208.20&1.50&1.50&2.44&213.15&0.00&0.00&4.66&259.06&19.36&19.36\\
1.54&212.53&1.50&1.50&2.46&203.04&0.00&0.00&5.09&213.22&70.44&70.44\\
1.61&207.16&1.50&1.50&2.47&200.07&0.00&0.00&10.49&170.37&6.93&6.93\\
1.64&209.15&1.50&1.50&2.50&202.09&1.04&1.04&16.63&157.89&19.57&19.57\\
1.71&253.14&1.69&1.69&2.57&270.52&0.66&0.66&27.57&143.95&29.49&29.49\\
\end{tabular}
\label{Tab.Data}
\end{table*}

\twocolumngrid
\normalsize

\begin{table}[H]
\caption{Disk and Bulge Values}\label{Tab.Data2}
\begin{center}
\begin{tabular}{ccc}
\specialrule{.1em}{.05em}{.05em}
\multicolumn{3}{c}{Disk}\\
\specialrule{.1em}{.05em}{.05em}
$N$&$M_0$&$\beta_d$\\
($10^{10}$)&($10^{30}kg$)&(${10}^{17}km$)\\
\hline
$6.5$&$1.988$&$1.08$\\\hline\\\\
\specialrule{.1em}{.05em}{.05em}
\multicolumn{3}{c}{Bulge}\\
\specialrule{.1em}{.05em}{.05em}
$\Sigma_{be}$&$\beta_b$&$\kappa$\\
($10^6kg\;km^{-2}$)&(${10}^{16}km$)&\\
\hline
$6.68$&$1.543$&$7.66945$\\
\hline
\end{tabular}
\end{center}
\end{table}
The velocity profile presented in Eq.(\ref{eq.all_vect_vel}) can now be used to find suitable values of the coupling parameter, $\alpha$. We do this by employing a least-squares approach to minimize the difference between the predicted and observed values. This is performed by setting  ${v(R_m)_{e_{{}_{\alpha{b}}}}}^2 = \alpha{u(R_m)_{e_{{}_{\alpha{b}}}}}^2$ and ${v(R_m)_{e_{{}_{\alpha{d}}}}}^2 = \alpha{u(R_m)_{e_{{}_{\alpha{d}}}}}^2$ and using the relation
\begin{equation}\label{eq fitting}
\alpha\approx\dfrac{1}{D_{max}}\displaystyle\sum_{m={0}}^{D_{max}-1}\frac{v_{m}^2-\left({v_{e_{{}_{\scaleto{TEGRd}{2pt}}}}}^2+{v_{e_{{}_{\scaleto{TEGRb}{2pt}}}}}^2\right)}{\left({{u_{e_{{}_{\alpha{b}}}}}^2}+{{u_{e_{{}_{\alpha{b}}}}}^2}\right)}.
\end{equation}
Here $D_{max}$ represents the size of the data set, i.e. the number of velocity points considered. \medskip

In Fig.(\ref{fig.grid}) we present plots for a range of values of $n$ with best fits for $\alpha$; this ranges from $n=-4$ to $n=1.4$. The corresponding GR plot is also being shown. Each plot consists of data points with associated error bars, the GR prediction for the Milky Way galaxy, and the best fit curve for the $f(T)$ model being considered where the coupling parameter $\alpha$ is being fitted. On a similar note, every plot has an embedded figure showing the difference between the best fit and the GR result. As both predicted curves plateau so does their associated difference in the embedded figure.  \medskip

\begin{figure*}[t]
\includegraphics[width=14cm]{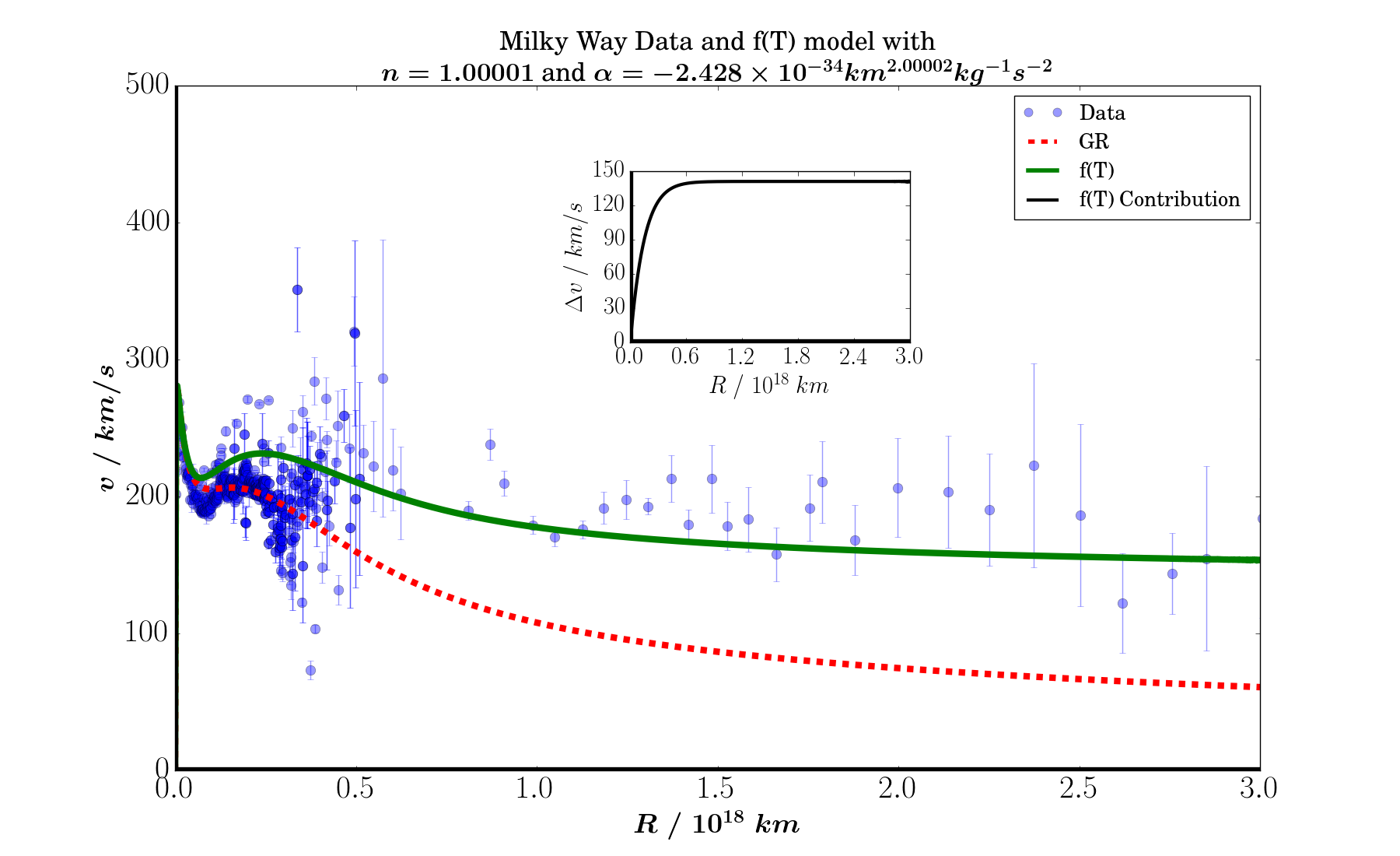}
\caption{Best fit model for the rotation profile in $f(T)$ gravity for the Milky Way galaxy with $n=1.00001$ and fit coupling parameter $\alpha=-2.428^{+1.303}_{-1.577}\times10^{-34}km^{2.00002}kg^{-1}s^{-2}$.}\label{fig.best.fit}
\end{figure*}

The first row shows negative values of $n$ where the behavior of the best fits are fairly similar in that they quickly diverge for data points in the disk region. Moreover, the predicted curves are completely nonphysical. In the following row, the special case of $n=0$ is considered. This corresponds to Einstein's GR with a constant similar to the cosmological constant. However, in this case the constant is designed to account for the anomalous rotation curves of galaxies. Naturally, the constant gives a near linear increase in velocity against the galactic radius. Again, the resulting velocity curve is unrealistic. The linear behavior is best seen in the embedded figure. As with the instances of negative $n$, the behavior performs worse when compared with GR. \medskip

Next we consider the $0<n<1$ range. The predictions from the $f(T)$ functional model become much more promising in this region as compared with the preceding values. As with the previous cases, the behavior in the core of the galaxy is relatively well behaved. Although as larger radial values are considered, we find that the predicted curve overshoots the velocity curve data points. The predicted curves get better as the value of $n$ approaches unity. However, this value of $n$ was excluded from the solution presented in Eq.(\ref{eq_Matteos_metric}) \cite{Matteo2015}. In the limit of $n$ actually taking on the unity value, the Lagrangian tends to a re-scaling of GR. \medskip

Finally the $1<n<\frac{3}{2}$ range is investigated against the Milky Way velocity curve. These provide the best behavior of graphs from all the ranges considered in this work. As in the other plots, a cut-off value for the maximum radius considered is set to $R=3\times10^{18}$. The predicted velocity profile does not have much interest for us beyond that point. The basic result from this range of values of $n$ is that as $n$ gets closer to unity the resulting velocity curve profiles behave better as compared to the observational points in question. \medskip

The best behaved value of index of the $f(T)$ Lagrangian is given by $n=1.00001$ which results in the best fit coupling parameter $\alpha=-2.428^{+1.303}_{-1.577}\times10^{-34}\text{km}^{2.00002}\text{kg}^{-1}\text{s}^{-2}$. This is shown in Fig.(\ref{fig.best.fit}). The resulting curve agrees with GR in the core part of the galaxy with slight variations in the central region. The difference between the two curves then plateaus, with the $f(T)$ function behaving better against observation than GR without dark matter contributions.

\twocolumngrid
\normalsize

\subsection{B. The three parameter fitting}

Heaving determined that the best range of values for $n$ is $1<n<\frac{3}{2}$, we now move to the next stage of the process which is to produce a three parameter fitting for the $f(T)$ rotation curve. This method should fit the surface mass density of the bulge $\Sigma_{be}$, the Mass of the Disk $M$ and the coupling constant $\alpha$. In this case, the fitting function proves to be nonlinear and as such an iterative method accompanied by a cross validation fitting is employed to determine these three parameters. \medskip

The results of this fit are presented in Table \ref{Tab.Data3} where the first column consists of the tested values of $n$, the second column is the minimized sum of the square of the residuals for the fitting function used, $S$, the third and fourth columns are the fits for the surface mass density of the galactic bulge, $\Sigma_{be}$, and the mass of the disk, $M$, respectively and the final column is the fit for the coupling constant $\alpha$. The resulting plots from these values can be seen in Fig.(\ref{fig.grid2}). From the plots it is possible to narrow down the best fit to two values of $n$, namely $n=1.00001$ and $n=1.000001$. The choice between these two instances can then be made using the second column of Table \ref{Tab.Data3}. The $S$ values for $n=1.00001$ and for $n=1.000001$ are $2.78\times10^9$ and $1.43\times10^{29}$ respectively. This clearly indicates that $n=1.00001$ provides us with the best $f(T)$ rotation curve fit. \medskip
\sisetup{exponent-base = 10,round-mode = figures, round-precision = 3,
  scientific-notation = true}
\renewcommand{\arraystretch}{1.5}
\begin{table}
\begin{center}
\caption{Fit of Bulge Surface Mass Density, Disk Mass and $f(T)$ Coupling Constant Values}\label{Tab.Data3}
\scriptsize
\hspace*{-.5cm}\begin{tabular}{ccccc}
\specialrule{.15em}{.05em}{.04em}
\multicolumn{1}{c}{$n$}&\multicolumn{1}{c}{$S$}&\multicolumn{1}{c}{$\Sigma_{b}$}&\multicolumn{1}{c}{$M$}&\multicolumn{1}{c}{$\alpha$}\\
\hline
\multicolumn{1}{c}{}&\multicolumn{1}{c}{}&\multicolumn{1}{c}{\small$10^6\;\text{kg}\;\text{km}^{-2}$}&\multicolumn{1}{c}{\small$10^{10}\;M_\odot$}&\multicolumn{1}{c}{\small$\text{km}^{2n}\text{kg}^{-1}\text{s}^{-2}$}\\
\specialrule{.15em}{.05em}{.04em}
1.4&\num{4.43867537192E+028}&\num{5.7822148832}&3.30&\num{-1.02930830332E-25}\\
1.3&\num{12382762139.9}&\num{5.664938424}&7.62&\num{-7.68217426605E-31}\\
1.2&\num{7687558949.72}&\num{5.5596150115}&3.04&\num{-1.4690655274E-31}\\
1.1&\num{7868393494.1}&\num{6.2143085554}&2.26&\num{-2.69050276391E-34}\\
1.01&\num{6750572196.48}&\num{5.9736212668}&7.49&\num{-1.73705610047E-47}\\
1.001&\num{8384778975.37}&\num{5.5143183084}&7.43&\num{-1.07550291076E-45}\\
1.0001&\num{24494042456.7}&\num{5.7628789052}&7.21&\num{-1.15357454546E-45}\\
1.00001&\num{2780771955.96}&\num{5.939846102}&4.08&\num{-4.50573152405E-34}\\
1.000001&\num{1.43462000896E+029}&\num{6.4898742539}&3.56&\num{-5.81567241159E-33}\\
\specialrule{.15em}{.05em}{.04em}
\end{tabular}
\end{center}
\end{table}
\normalsize
The best fit value for $n$ results in a bulge surface mass density of $5.94\times10^6\;\text{kg}\;\text{km}^{-2}$, a disk mass of $4.08\times10^{10}M_\odot$ and a coupling constant of \num{-4.50573152405E-34}$\;\text{km}^{2n}\;\text{kg}^{-1}\;\text{s}^{-2}$. The Surface mass density of the bulge $\Sigma_{\text{be}}$ is directly related to the bulge mass through the equation 
\begin{equation}\label{denstoM}
M(r)=4\int_0^R{\displaystyle{\int_r^{\infty}}\dfrac{d\Sigma_b(x)}{dx}\dfrac{r^2}{\sqrt{x^2-r^2}}{r^2}dxdr},
\end{equation}
where 
\begin{equation}\label{de_Vaucouleurs_profile2}
\Sigma_{\text{b}}(x)=\Sigma_{\text{be}} \text{Exp}\left[-\kappa\left(\left(\frac{x}{\beta_b}\right)^{1/4}-1\right)\right].
\end{equation}
Given the fit values for $\Sigma_{\text{be}}$ and using Eq.(\ref{denstoM}) and Eq.(\ref{de_Vaucouleurs_profile2}), the total mass of the bulge can be calculated to be $1.2\times10^{10}M_\odot$. Thus, this gives a total galactic luminous matter mass of $5.36\times10^{10}M_\odot$ for the Milky Way.\medskip

\onecolumngrid
\begin{figure}[H]
\hspace{-1.3cm}\begin{minipage}{\textwidth}
\setlength{\tabcolsep}{-9.5pt} 
\renewcommand{\arraystretch}{-.5}
\begin{tabular}{ccc}
\includegraphics[width=65mm]{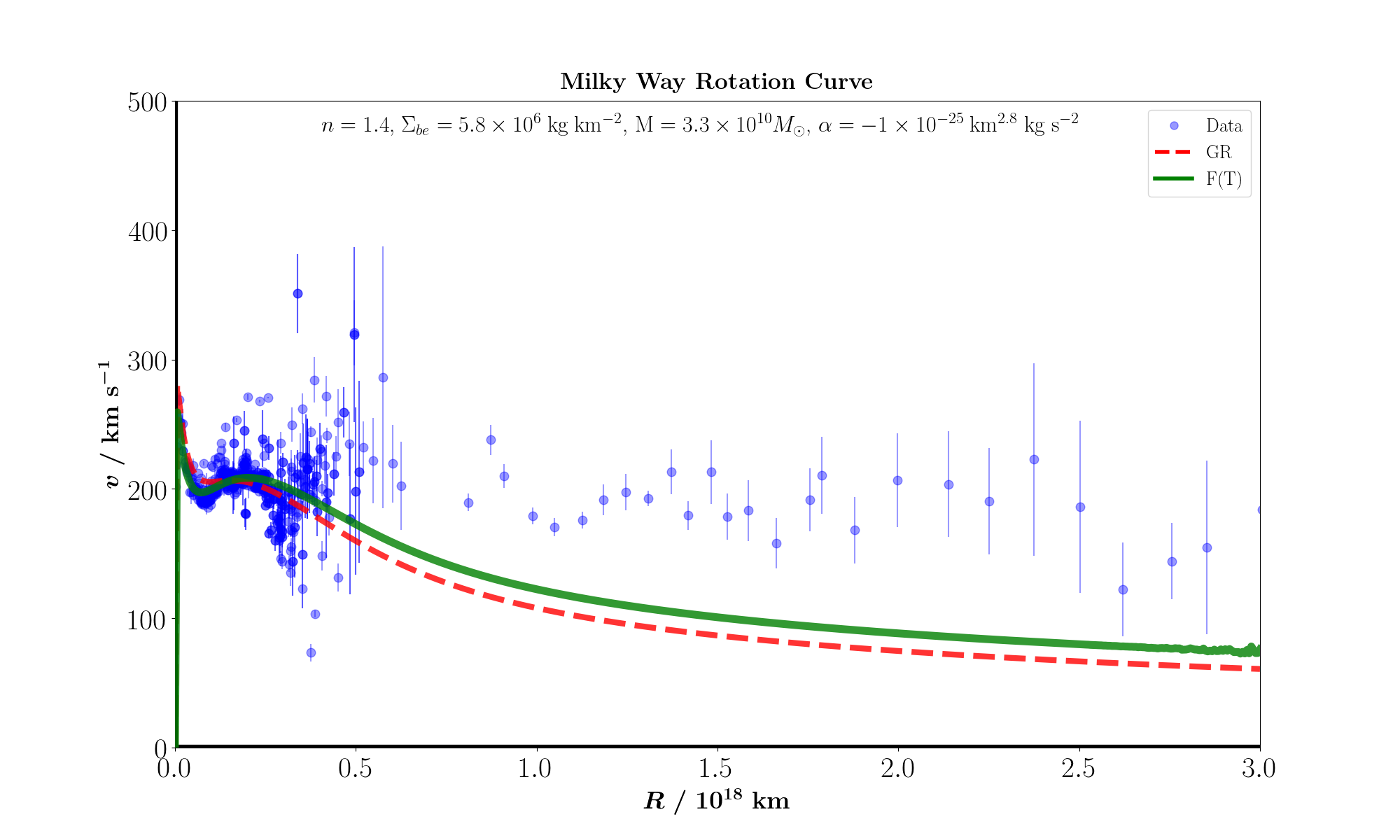}&
\includegraphics[width=65mm]{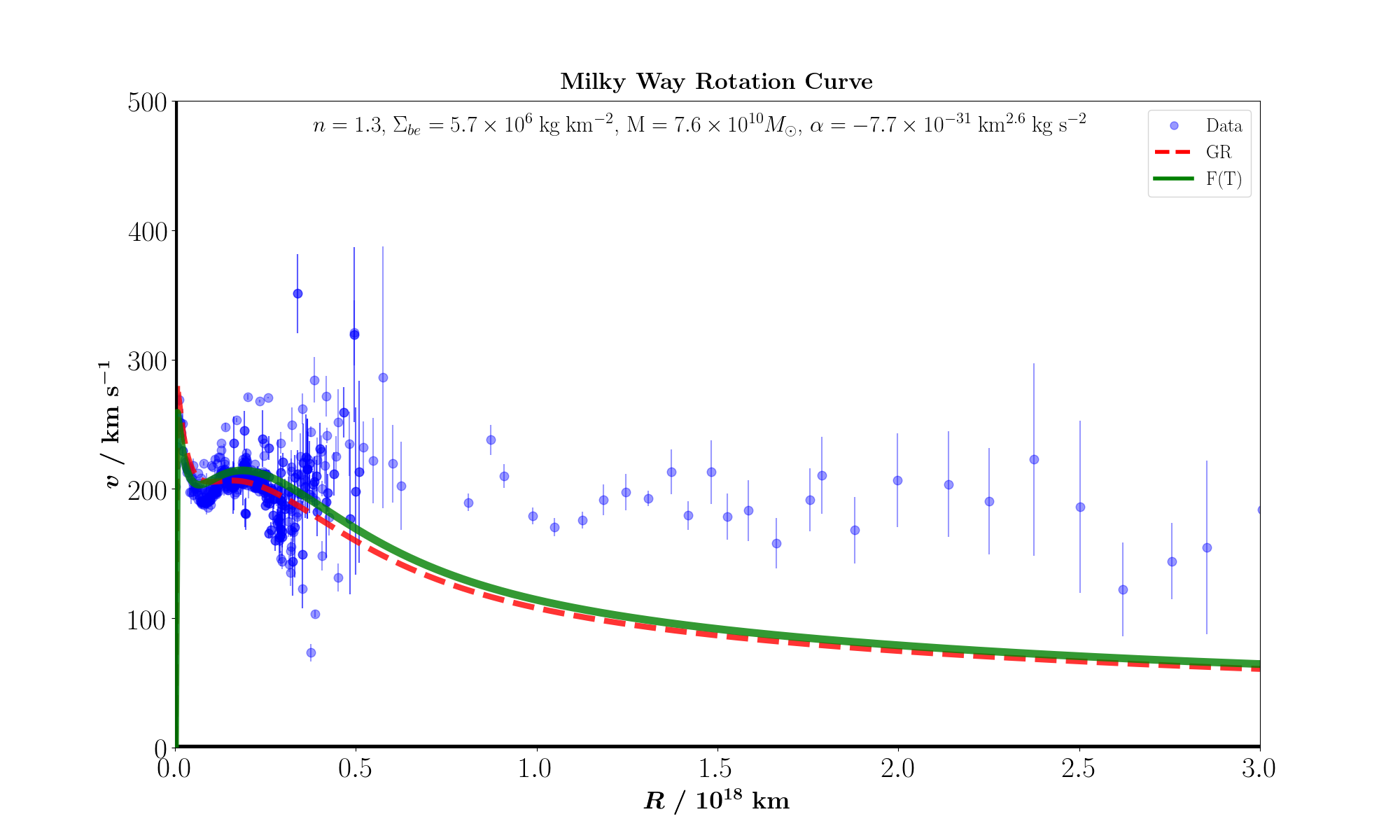}&   
\includegraphics[width=65mm]{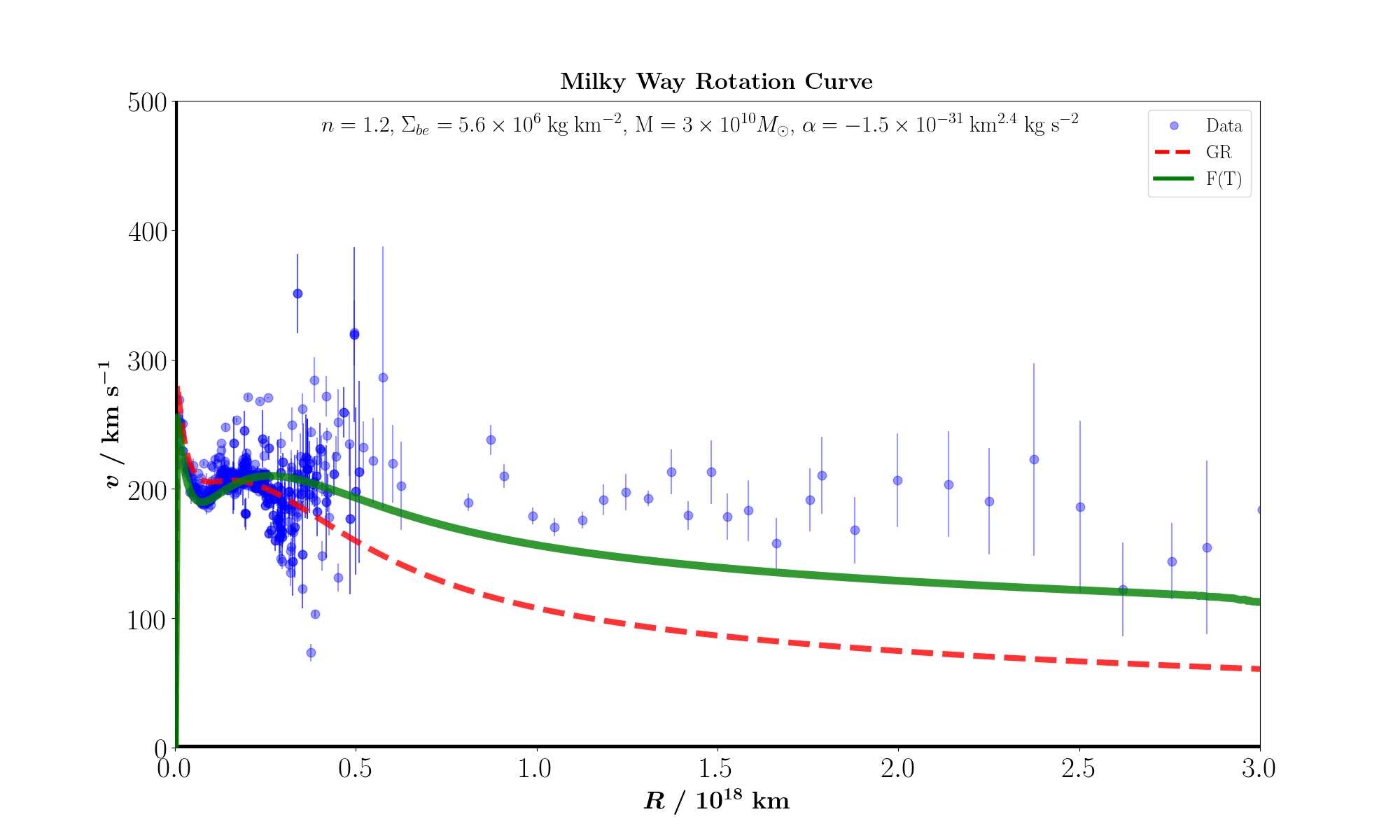}\\
\includegraphics[width=65mm]{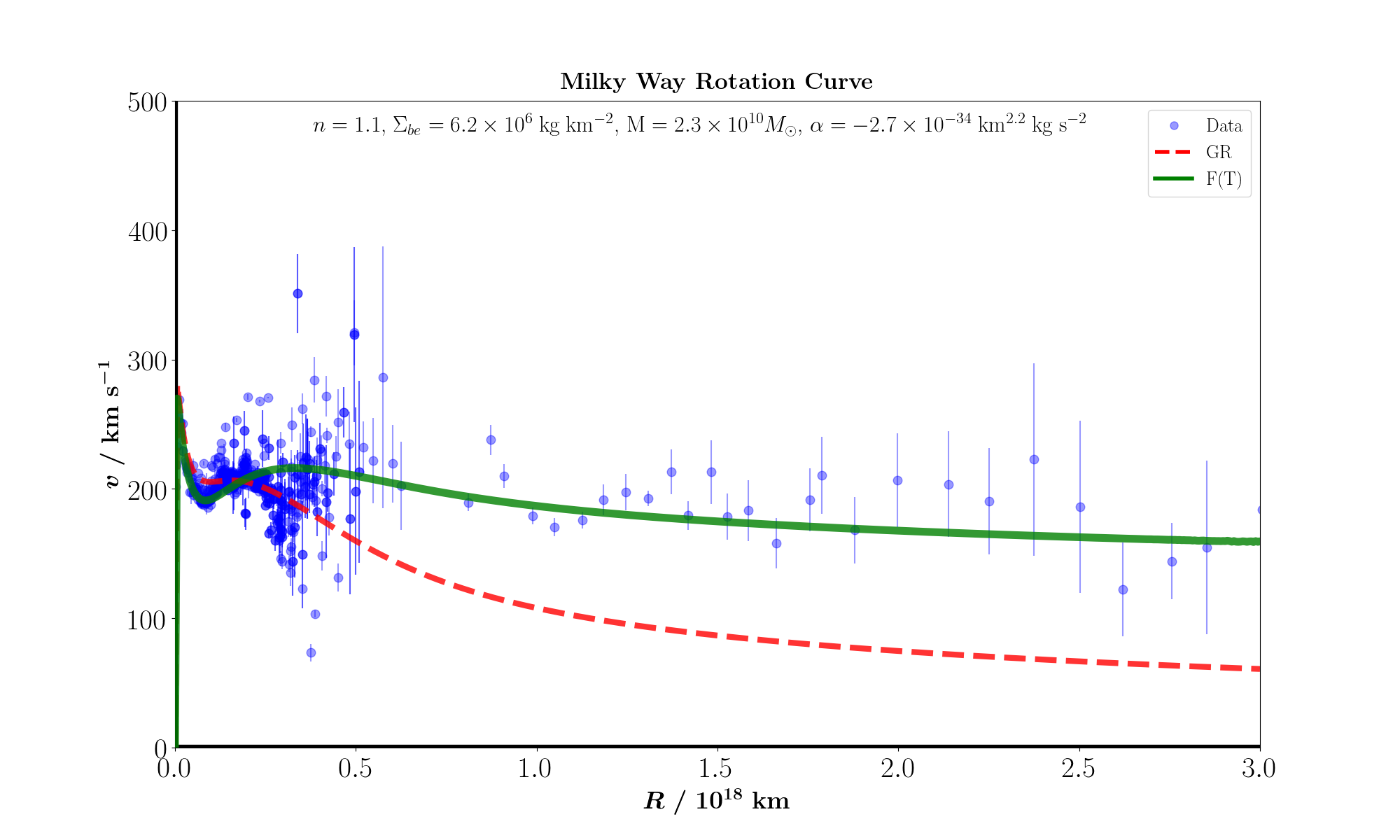}&
\includegraphics[width=65mm]{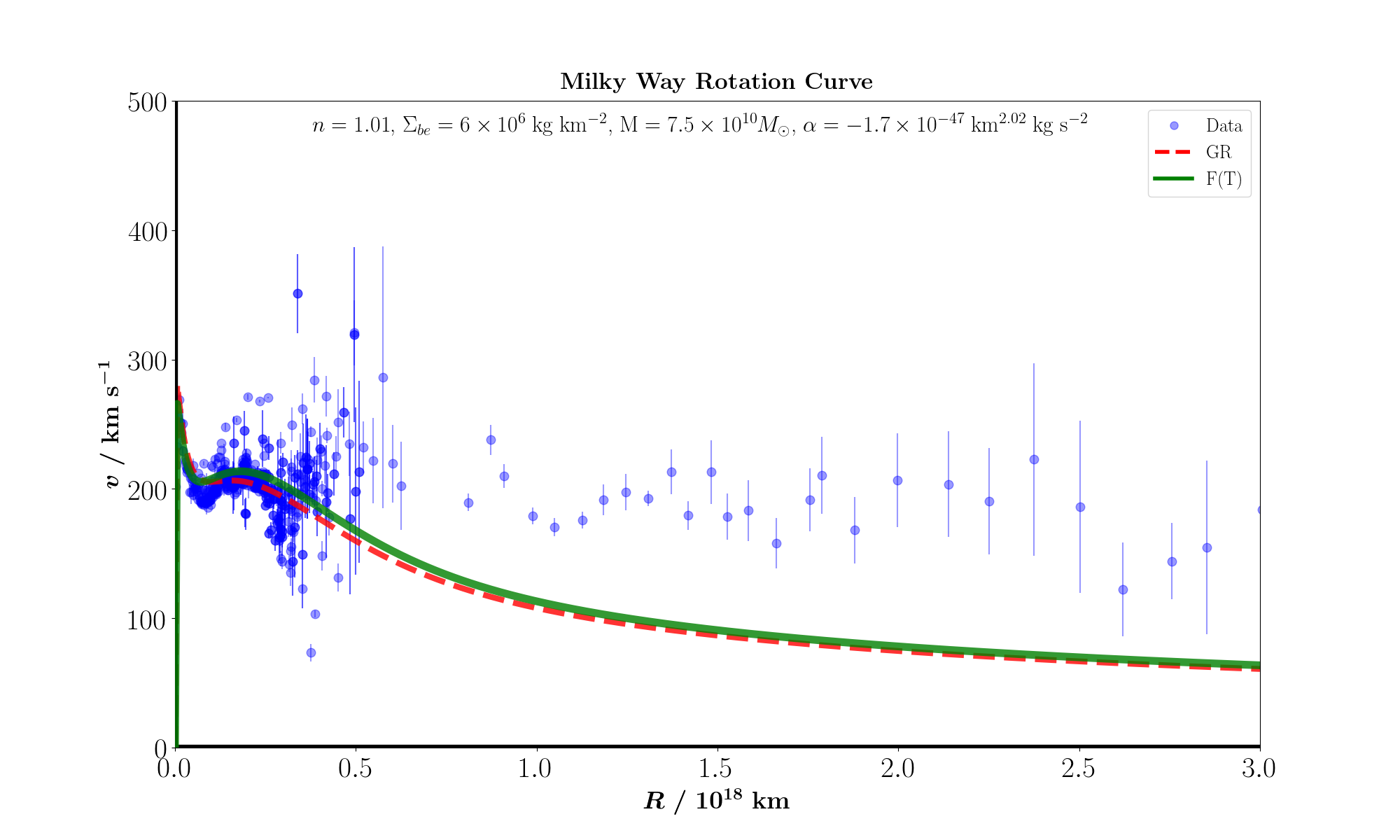}&
\includegraphics[width=65mm]{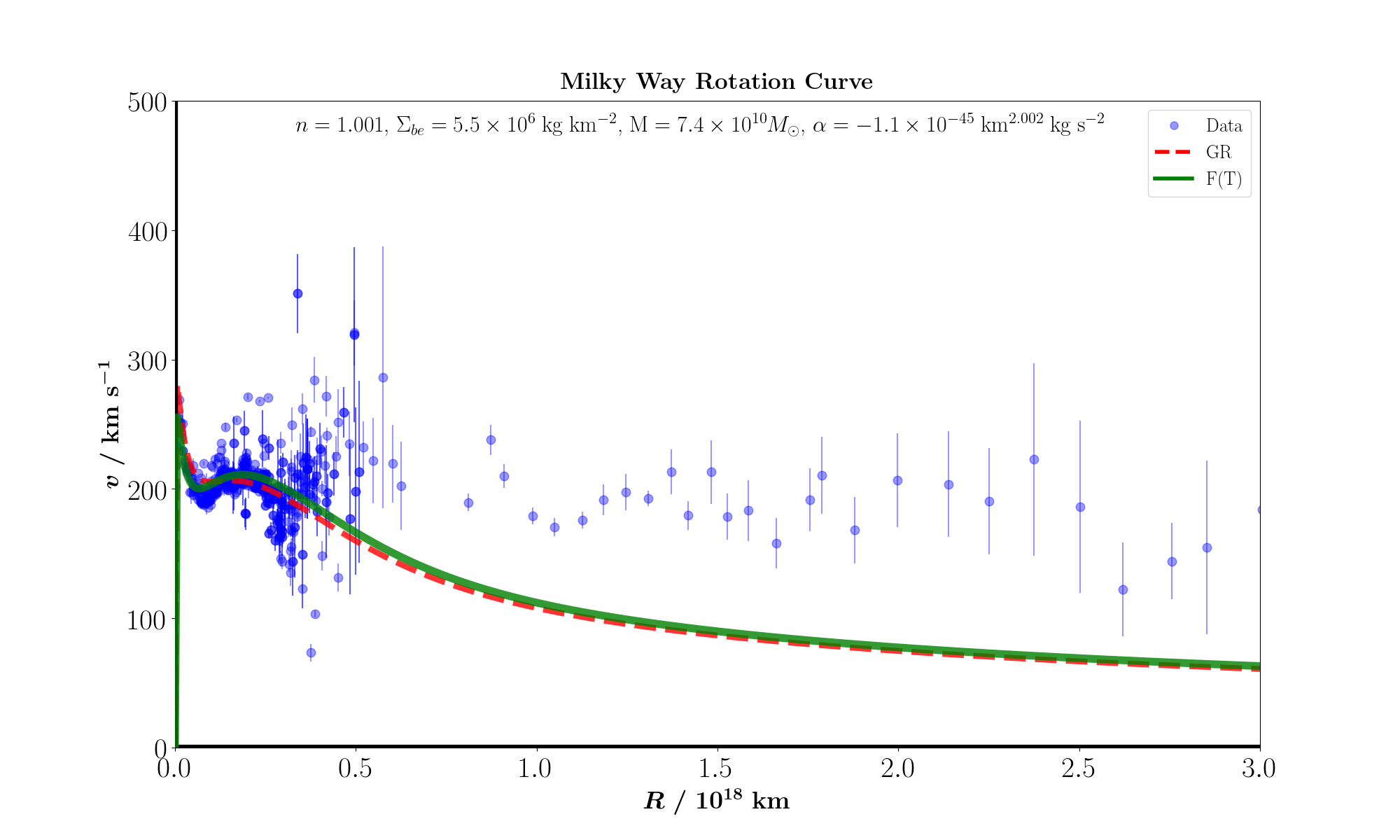}\\
\includegraphics[width=65mm]{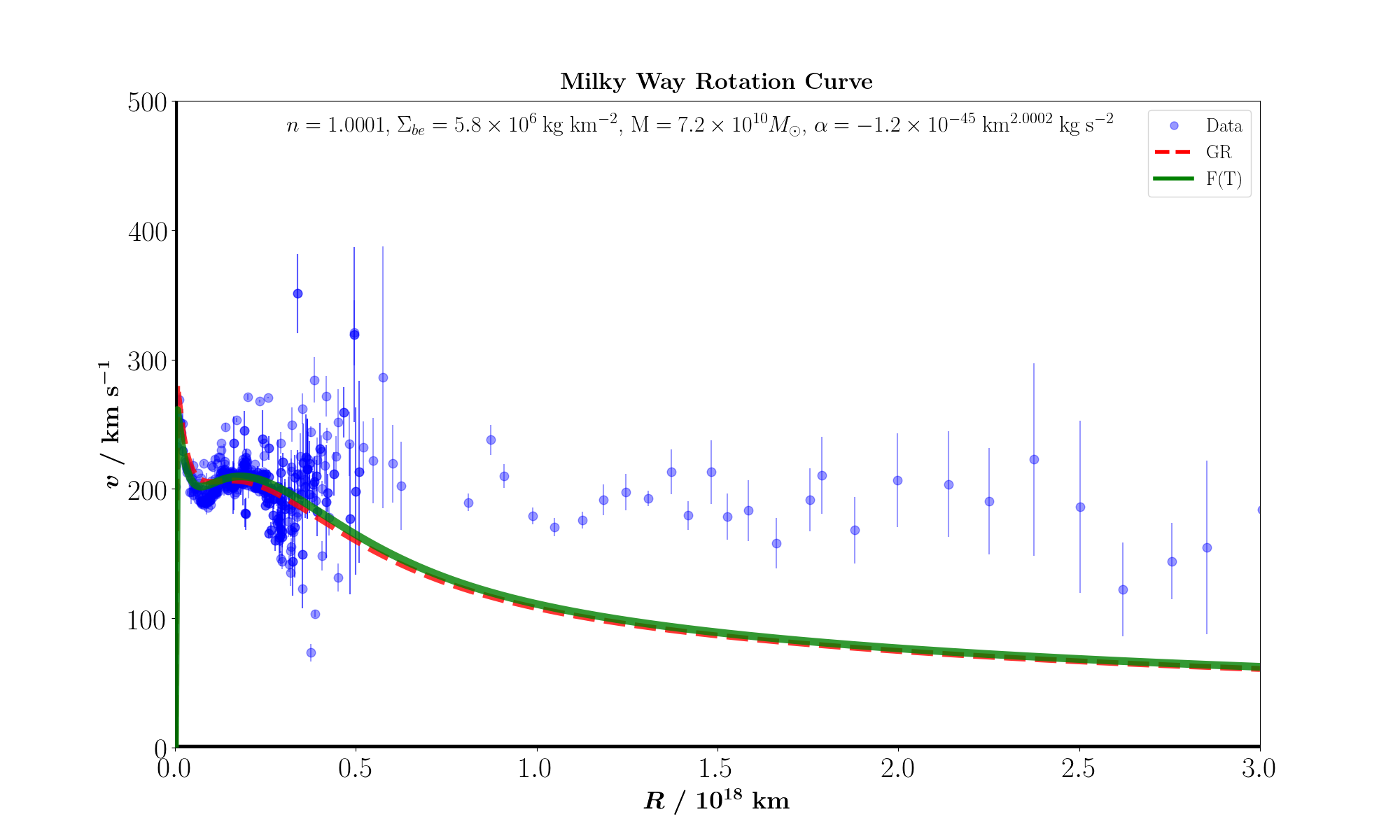}&
\includegraphics[width=65mm]{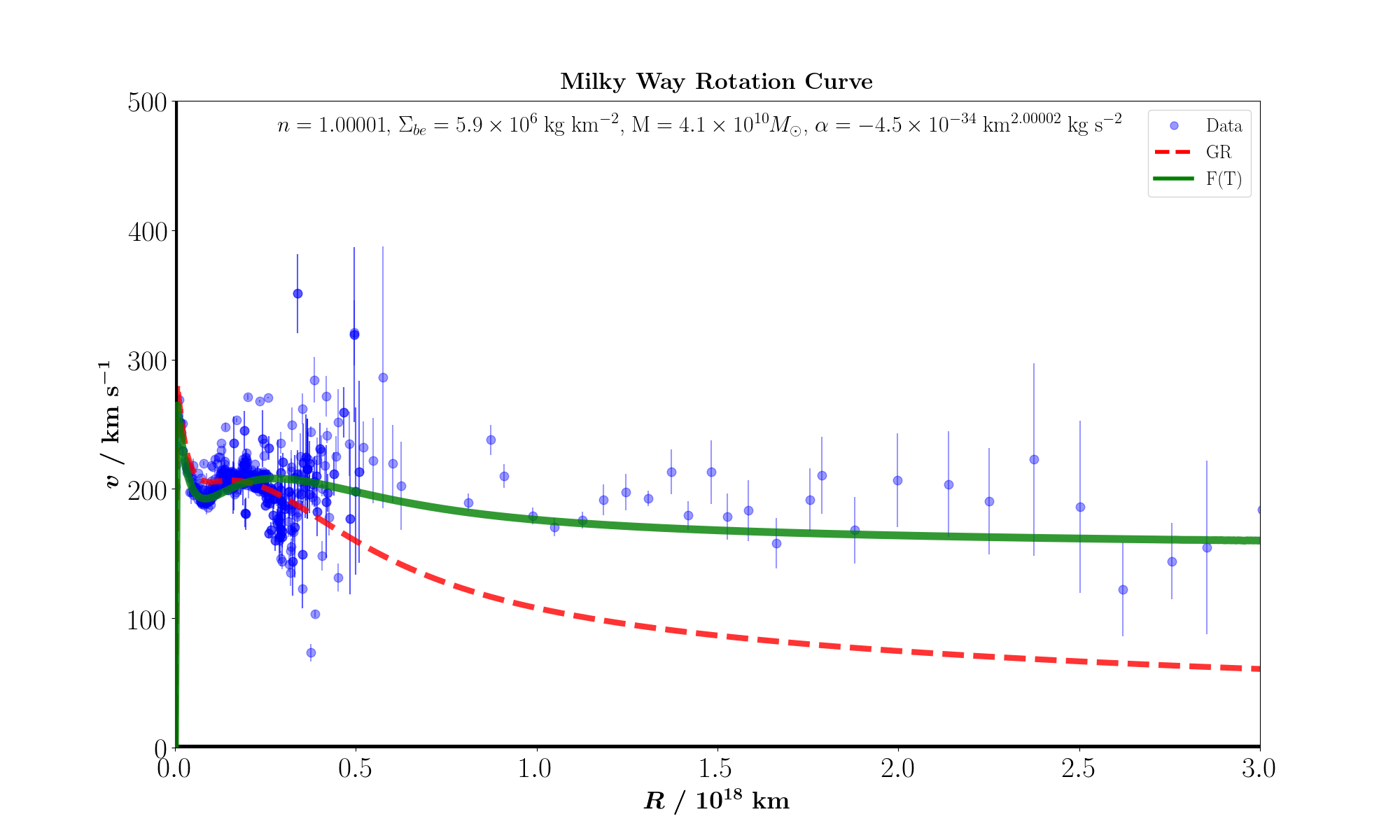}&
\includegraphics[width=65mm]{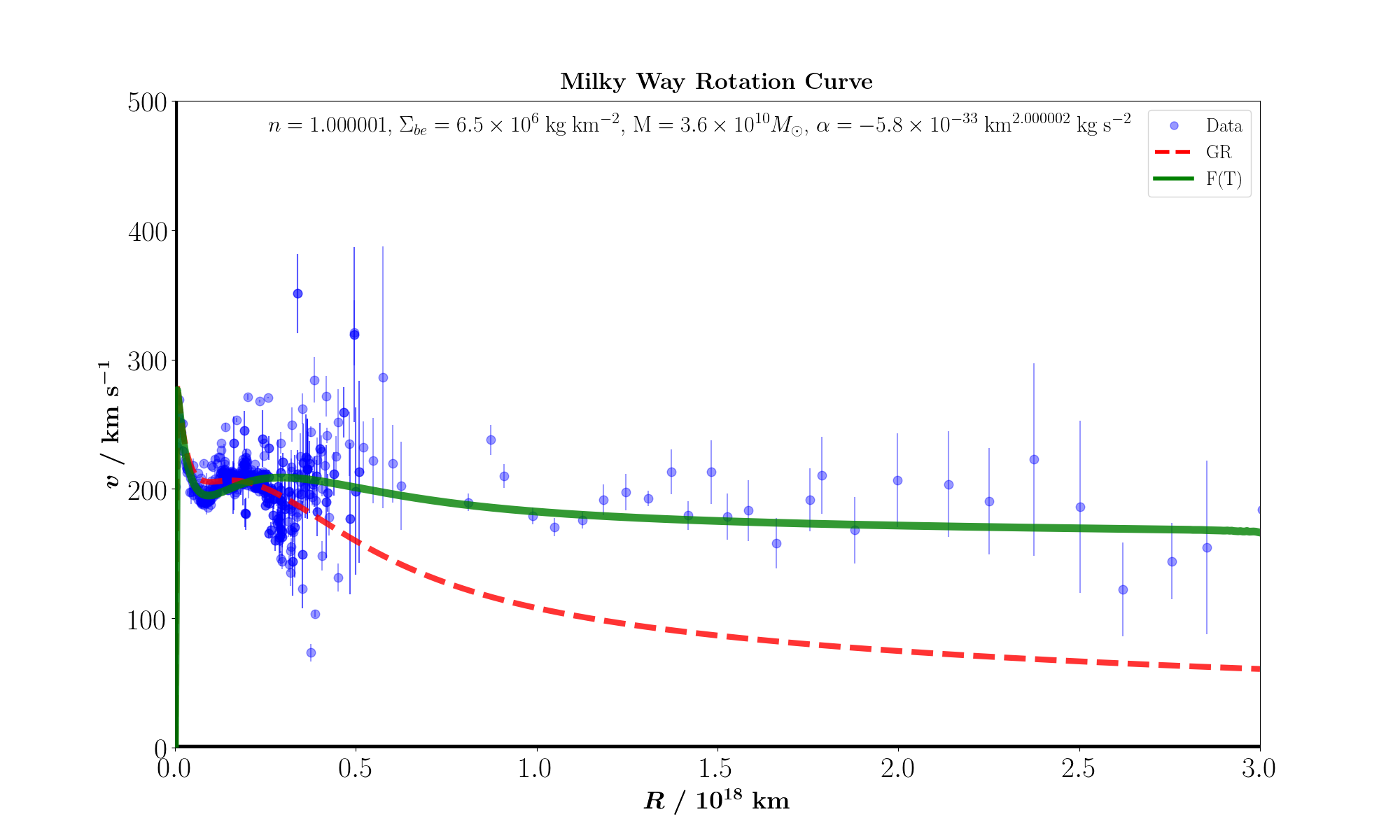}\\
\end{tabular}
\end{minipage}
\parbox{\textwidth}{\caption{$f(T)$ gravity rotational velocities with $n$ values ranging from $1.4$ to $1.000001$, with a fitted bulge surface mass density $\Sigma_{be}$, disk mass $M$ and coupling constant $\alpha$.}\label{fig.grid2}}
\end{figure}


\begin{figure*}[t]
\includegraphics[width=14cm]{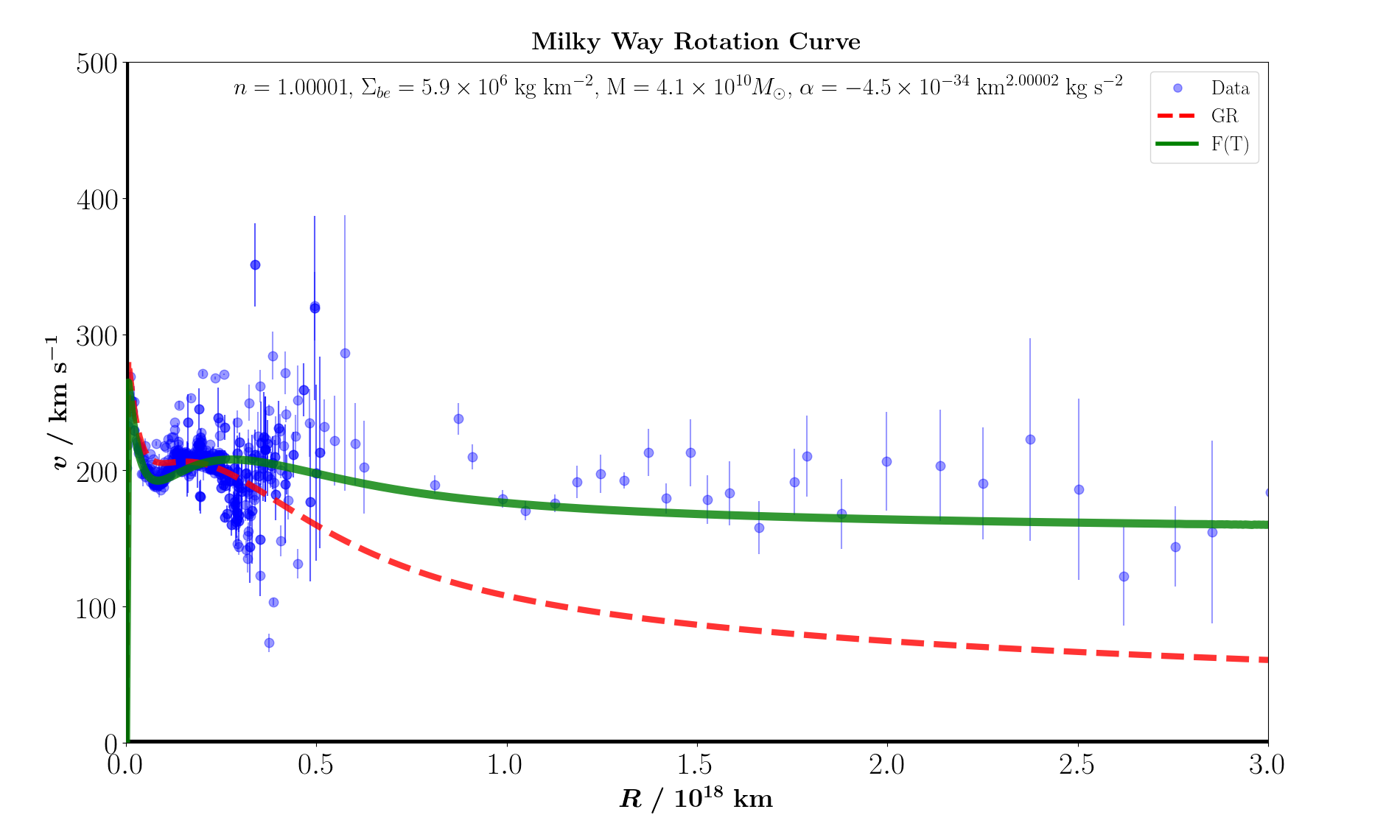}
\caption{Best fit model for the rotation profile in $f(T)$ gravity for the Milky Way galaxy with $n=1.00001$ and fit coupling parameter $\alpha=$\num{-4.50573152405E-34}$\;\text{km}^{2n}\;\text{kg}^{-1}\;\text{s}^{-2}$, bulge surface mass density of $5.94\times10^6\;\text{kg}\;\text{km}^{-2}$ and disk mass $4.08\times10^{10}M_\odot$.}\label{fig.best.fit2}
\end{figure*}

\twocolumngrid
\normalsize

We now compare this value to known values for the mass of the Milky Way without dark matter. Ref.\cite{Mass2} reports a total combined mass of $5.22\times10^{11}M_\odot$ of the Milky Way which includes the dark matter contribution. In Ref.\cite{Mass1}, it is given that the bulge mass is $0.91\pm0.07\times10^{10}M_\odot$ and the disk mass is $5.17\pm1.11\times10^{10}M_\odot$  giving a total mass of $6.08\pm1.14\times10^{10}M_\odot$. The total mass of these contributions to the galaxy obtained in this work agrees with the order of the total masses from both Ref.\cite{Mass1} and Ref.\cite{Mass2}. Furthermore, the values of the total mass, and the disk mass obtained here even fall within the error bars of the respective masses given by Ref.\cite{Mass1}. \medskip

Finally we compare this fit, which for ease of reference is given in Fig.(\ref{fig.best.fit2}), with the best fit while just fitting for $\alpha$, Fig.(\ref{fig.best.fit}). With this new three parameter fitting the $f(T)$ rotation curve successfully interprets the rotational data profile at the inner parts of the galaxy while still retaining the plateau effect obtained in the first fit. Having a value of $n$ that is greater than $1$ also prevents the possibility that the velocity curve will tend to infinity at larger radii, which would lead to an unrealistic prediction. \medskip

\section{V. What about other galaxy morphologies?}

In this section we consider three other galaxies apart from the Milky way which adhere to separate galaxy morphologies. Specifically we consider the bright spiral galaxy NGC 3198, the low surface brightness galaxy UGC 128 and the irregular dwarf galaxy DDO 154. These were chosen so as to test the fit on different types of galaxies since these clearly exhibit their own separate morphologies. \medskip

The rotation curve data for these galaxies is shown in Table \ref{Tab.Data4}. The data used for NGC 3198 and UGC 128 was obtained from Ref.\cite{SPARC2016}, and the data for DDO154 was obtained from  Refs.\cite{SPARC2016,1998DDO154}.\medskip

\begin{center}
\onecolumngrid
\sisetup{exponent-base = 10,round-mode = figures, round-precision = 3,
  scientific-notation = true}
\setlength{\tabcolsep}{6pt}
\renewcommand{\arraystretch}{1.2}
\begin{table}[H]
\begin{center}
\parbox{\textwidth}{\caption{Best fit of Galactic Mass for NGC 3198, UGC 128 and DDO 154}\label{Tab.Data5}}
\hspace*{1cm}\begin{tabular}{lccccccc}
\specialrule{.09em}{.05em}{.04em}
\multicolumn{1}{c}{$Name$}&\multicolumn{1}{c}{$\beta_d$}&\multicolumn{1}{c}{$M_{\text{GR}}$}&\multicolumn{1}{c}{$n$}&\multicolumn{1}{c}{$M$}&\multicolumn{1}{c}{$\alpha$}&\multicolumn{2}{c}{$Refs$}\\
\hline
\multicolumn{1}{c}{}&\multicolumn{1}{c}{$10^{16}\;$km}&\multicolumn{1}{c}{$10^{10}\;M_\odot$}&\multicolumn{1}{c}{}&\multicolumn{1}{c}{$10^{10}\;M_\odot$}&\multicolumn{1}{c}{$10^{-34}\;\text{km}^{2n}kg^{-1}\text{s}^{-2}$}&\multicolumn{1}{c}{$\beta_d$}&\multicolumn{1}{c}{$M_{\text{GR}}$}\\
\specialrule{.1em}{.05em}{.04em}
NGC3198	&	9.69	&	4.500	&	1.00001	&	2.49	&	-4.51	&	\cite{SPARC2016}	&	\cite{NGCMass}	\\
UGC128	&	18.36	&	4.300	&	1.00001	&	2.47	&	-4.51	&	\cite{SPARC2016}	&	\cite{UGCMass}	\\
DDO154	&	1.23	&	0.003	&	1.00001	&	0.15	&	-4.51	&	\cite{DDOMass}	&	\cite{DDOMass}	\\
\specialrule{.09em}{.05em}{.04em}
\end{tabular}
\end{center}
\end{table}
\end{center}
\twocolumngrid
\normalsize

\begin{figure}[H]
\begin{minipage}{\textwidth}
\setlength{\tabcolsep}{-9.6pt} 
\begin{tabular}{c}
\includegraphics[width=79mm]{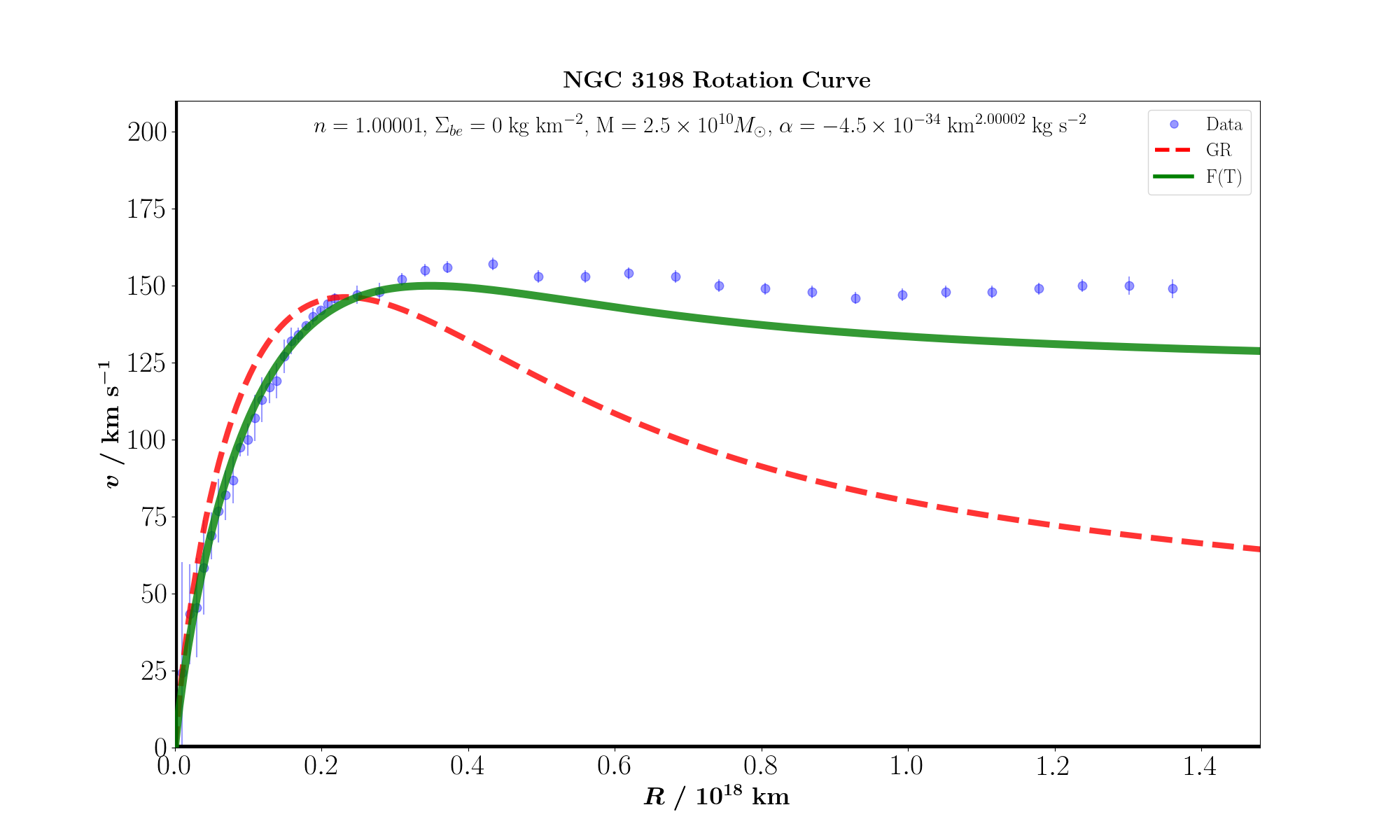}\label{galaxy_one}\\   
\includegraphics[width=79mm]{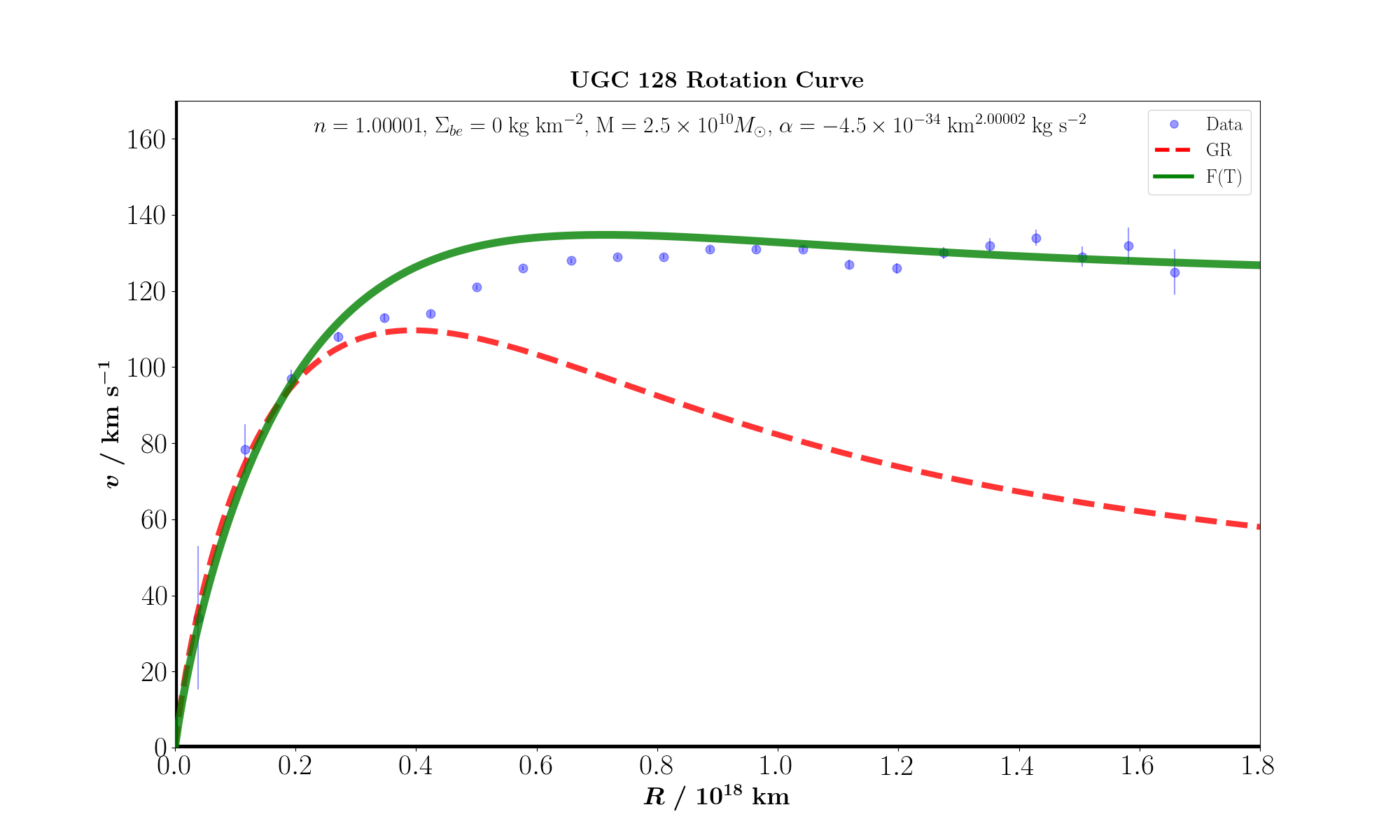}\label{galaxy_two}\\
\includegraphics[width=79mm]{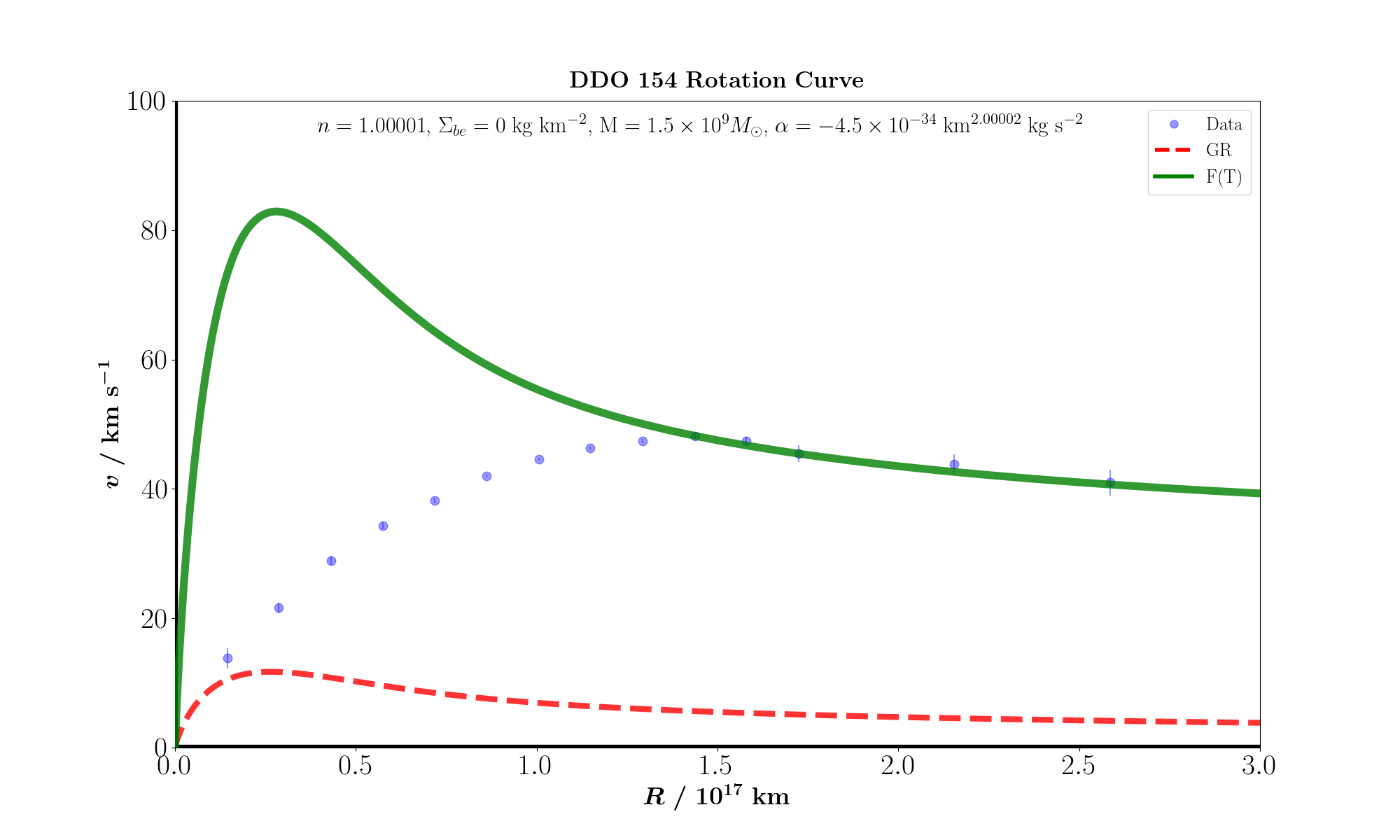}\label{galaxy_three}\\
\end{tabular}
\end{minipage}
\caption{In these plots the full green curve represents the $f(T)$ gravity rotational velocity curves with $n=1.00001$, with $\alpha=\num{-4.50573152405E-34}\;\text{km}^{2n}\;\text{kg}^{-1}\;\text{s}^{-2}$,  and with masses as shown in Table \ref{Tab.Data5}. The red dashed curve represent the general relativistic rotation curve and the data points are the rotational velocity data sets provided in Table \ref{Tab.Data4}.}
\label{fig.grid3}
\end{figure}

In these three cases the surface mass density of the galaxies' bulges, $\Sigma_{be}$, is assumed to be zero as they are variants of disk galaxies. The mass of their disks, $M$, was fit while keeping the coupling constant $\alpha$ fixed at the value obtained by the best fit of the Milky Way in the previous section. The results for the fit can be seen in Table \ref{Tab.Data5}. The first column of Table \ref{Tab.Data5} represents the name of the galaxy considered, the second column gives the disk scale radii, $\beta_d$, the third column gives the galactic masses as predicted by GR, $M_{\text{GR}}$ and the fourth gives the value of the constant $n$ at which these galaxies were fit. The fifth and sixth columns provide the fitting values for the disk masses and the value of the coupling constant $\alpha$ considered respectively. Finally the last column provides the data sources for $\beta_d$ and $M_{\text{GR}}$. Using these results the three plots shown in Fig.(\ref{fig.grid3}) were generated. \medskip

The first galaxy considered, NGC 3198, is a bright spiral galaxy \cite{NGCMass}. The data available clearly shows that the rotational velocities plateau at around $150\;\text{km}\;\text{s}^{-1}$. Using the same fitting method used to fit the $f(T)$ rotation curve for the Milky Way results in a luminous matter disk mass of $2.49\times10^{10}\;M_\odot$. The mass obtained is of the expected order and is smaller than the mass predicted by GR. Overall the $f(T)$ fit is by far more accurate than the General Relativistic one. Up to a galactic radius of around $3\times10^{17}\;$km the rotation curve fits perfectly with the data, while further out the predicted curve falls below the velocities of the data points. The reason for this drop in the fit curve may be attributed to the fact that no spiral arm equation was included in the fit which could also be the partly the reason why the mass is somewhat less than expected. \medskip

The second galaxy considered was UGC 128, a low surface brightness disk galaxy \cite{UGC128desk}. Here the data points once again produce an evident plateau at around $140\;\text{km}\;\text{s}^{-1}$. Once again the order of the disk mass is as expected though the magnitude is smaller than that predicted by the GR values. It should be noted that in both this case and the case of NGC 3198 the masses where fit while also fitting for the mass of non-luminous matter. As such, the larger values can, at least in part, be attributed to a missing distribution of mass from non-luminous to luminous. From the plot in Fig.(\ref{fig.grid3}.2) we observe that the $f(T)$ fit is accurate throughout accept for a drop in the data velocity magnitudes between $3\times10^{17}$km and $7\times10^{17}$km. \medskip

The third and last galaxy considered, DDO 154, is an irregular dwarf galaxy. In this case it proved not possible to fit the $f(T)$ rotation curve to the data points. This can clearly be seen in the last plot presented in Fig.(\ref{fig.grid3}.3). The reason for this is that most of DDO 154's mass is dominated by a distribution of gasses \cite{DDOMass} that can neither be classified as being part of the disk nor a type of bulge. As such in order to correctly test this galaxy and others like it one would have to derive the correct velocity contributions of such gas distributions. The red dashed line is the GR disk contribution which also behaves much worse than the other galaxy morphology types.

\onecolumngrid
\newpage
\setlength{\tabcolsep}{4pt}
\sisetup{round-integer-to-decimal,round-mode = places, round-precision = 2,
  scientific-notation = false}
\renewcommand{\arraystretch}{1.3}
\begin{table}[H]
\caption{Galactic rotation curve data for NGC 3198, UGC 128 and DDO 154. Column (1): radial distance from the galactic center. Column (2): velocity at that radial position. Column (3): error bar value.}\label{Tab.Data4}
\begin{center}
\hspace*{-.4cm}\begin{tabular}{lccclccclccc}
\specialrule{.09em}{.05em}{.04em}
\multicolumn{1}{c}{Name}&\multicolumn{1}{c}{$R$}&\multicolumn{1}{c}{$v$}&\multicolumn{1}{c}{$e_{v}$}&\multicolumn{1}{c}{Name}&\multicolumn{1}{c}{$R$}&\multicolumn{1}{c}{$v$}&\multicolumn{1}{c}{$e_{v}$}&\multicolumn{1}{c}{Name}&\multicolumn{1}{c}{$R$}&\multicolumn{1}{c}{$v$}&\multicolumn{1}{c}{$e_{v}$}\\
\hline
\multicolumn{1}{c}{}&\multicolumn{1}{c}{$10^{16}\text{km}$}&\multicolumn{1}{c}{$\text{km}\;\text{s}^{-1}$}&\multicolumn{1}{c}{$\text{km}\;\text{s}^{-1}$}&\multicolumn{1}{c}{}&\multicolumn{1}{c}{$10^{16}\text{km}$}&\multicolumn{1}{c}{$\text{km}\;\text{s}^{-1}$}&\multicolumn{1}{c}{$\text{km}\;\text{s}^{-1}$}&\multicolumn{1}{c}{}&\multicolumn{1}{c}{$10^{16}\text{km}$}&\multicolumn{1}{c}{$\text{km}\;\text{s}^{-1}$}&\multicolumn{1}{c}{$\text{km}\;\text{s}^{-1}$}\\
\specialrule{.1em}{.05em}{.04em}
NGC3198&\num{0.987424}&\num{24.4}&\num{35.9}&UGC128&\num{3.857125}&\num{34.1}&\num{18.8}&DDO154&\num{1.450279}&\num{13.8}&\num{1.6}\\
NGC3198&\num{1.974848}&\num{43.3}&\num{16.3}&UGC128&\num{11.571375}&\num{78.4}&\num{6.54}&DDO154&\num{1.511993}&\num{13.8}&\num{1.6}\\
NGC3198&\num{2.962272}&\num{45.5}&\num{16.1}&UGC128&\num{19.285625}&\num{96.9}&\num{2.44}&DDO154&\num{2.869701}&\num{21.6}&\num{0.8}\\
NGC3198&\num{3.949696}&\num{58.5}&\num{15.4}&UGC128&\num{27.030732}&\num{108}&\num{1.26}&DDO154&\num{3.054843}&\num{21.6}&\num{0.8}\\
NGC3198&\num{4.967977}&\num{68.8}&\num{7.61}&UGC128&\num{34.714125}&\num{113}&\num{1}&DDO154&\num{4.31998}&\num{28.9}&\num{0.7}\\
NGC3198&\num{5.955401}&\num{76.9}&\num{10.3}&UGC128&\num{42.397518}&\num{114}&\num{0.86}&DDO154&\num{4.566836}&\num{28.9}&\num{0.7}\\
NGC3198&\num{6.911968}&\num{82}&\num{8.09}&UGC128&\num{50.050054}&\num{121}&\num{0.68}&DDO154&\num{5.739402}&\num{34.3}&\num{0.5}\\
NGC3198&\num{7.930249}&\num{86.9}&\num{7.6}&UGC128&\num{57.733447}&\num{126}&\num{0.55}&DDO154&\num{6.078829}&\num{34.3}&\num{0.5}\\
NGC3198&\num{8.917673}&\num{97.6}&\num{3.03}&UGC128&\num{65.72541}&\num{128}&\num{0.63}&DDO154&\num{7.189681}&\num{38.2}&\num{0.4}\\
NGC3198&\num{9.905097}&\num{100}&\num{5.31}&UGC128&\num{73.408803}&\num{129}&\num{0.61}&DDO154&\num{7.621679}&\num{38.2}&\num{0.4}\\
NGC3198&\num{10.923378}&\num{107}&\num{7.51}&UGC128&\num{81.092196}&\num{129}&\num{0.65}&DDO154&\num{8.609103}&\num{42}&\num{0.2}\\
NGC3198&\num{11.879945}&\num{113}&\num{7.32}&UGC128&\num{88.775589}&\num{131}&\num{0.73}&DDO154&\num{9.133672}&\num{42}&\num{0.2}\\
NGC3198&\num{12.867369}&\num{117}&\num{5.21}&UGC128&\num{96.428125}&\num{131}&\num{0.83}&DDO154&\num{10.059382}&\num{44.6}&\num{0.2}\\
NGC3198&\num{13.88565}&\num{119}&\num{5.67}&UGC128&\num{104.111518}&\num{131}&\num{0.75}&DDO154&\num{10.676522}&\num{44.6}&\num{0.2}\\
NGC3198&\num{14.873074}&\num{127}&\num{5.39}&UGC128&\num{111.794911}&\num{127}&\num{1.24}&DDO154&\num{11.478804}&\num{46.3}&\num{0.2}\\
NGC3198&\num{15.891355}&\num{132}&\num{4.34}&UGC128&\num{119.786874}&\num{126}&\num{1.4}&DDO154&\num{12.188515}&\num{46.3}&\num{0.2}\\
NGC3198&\num{16.847922}&\num{134}&\num{2.36}&UGC128&\num{127.470267}&\num{130}&\num{1.62}&DDO154&\num{12.929083}&\num{47.4}&\num{0.3}\\
NGC3198&\num{17.835346}&\num{137}&\num{0.89}&UGC128&\num{135.15366}&\num{132}&\num{2.01}&DDO154&\num{13.700508}&\num{47.4}&\num{0.3}\\
NGC3198&\num{18.82277}&\num{140}&\num{2.84}&UGC128&\num{142.806196}&\num{134}&\num{2.1}&DDO154&\num{14.379362}&\num{48.2}&\num{0.6}\\
NGC3198&\num{19.841051}&\num{142}&\num{0.88}&UGC128&\num{150.489589}&\num{129}&\num{2.66}&DDO154&\num{15.243358}&\num{48.2}&\num{0.6}\\
NGC3198&\num{20.797618}&\num{144}&\num{1.23}&UGC128&\num{158.172982}&\num{132}&\num{4.67}&DDO154&\num{15.798784}&\num{47.4}&\num{0.7}\\
NGC3198&\num{21.785042}&\num{146}&\num{1.57}&UGC128&\num{165.856375}&\num{125}&\num{5.95}&DDO154&\num{16.755351}&\num{47.4}&\num{0.7}\\
NGC3198&\num{24.809028}&\num{147}&\num{3}&\multicolumn{1}{c}{-}&\multicolumn{1}{c}{-}&\multicolumn{1}{c}{-}&\multicolumn{1}{c}{-}&DDO154&\num{17.249063}&\num{45.5}&\num{1.3}\\
NGC3198&\num{27.894728}&\num{148}&\num{3}&\multicolumn{1}{c}{-}&\multicolumn{1}{c}{-}&\multicolumn{1}{c}{-}&\multicolumn{1}{c}{-}&DDO154&\num{18.267344}&\num{45.5}&\num{1.3}\\
NGC3198&\num{30.980428}&\num{152}&\num{2}&\multicolumn{1}{c}{-}&\multicolumn{1}{c}{-}&\multicolumn{1}{c}{-}&\multicolumn{1}{c}{-}&DDO154&\num{21.538186}&\num{43.8}&\num{1.5}\\
NGC3198&\num{34.066128}&\num{155}&\num{2}&\multicolumn{1}{c}{-}&\multicolumn{1}{c}{-}&\multicolumn{1}{c}{-}&\multicolumn{1}{c}{-}&DDO154&\num{25.858166}&\num{41}&\num{2}\\
NGC3198&\num{37.182685}&\num{156}&\num{2}&\multicolumn{1}{c}{-}&\multicolumn{1}{c}{-}&\multicolumn{1}{c}{-}&\multicolumn{1}{c}{-}&\multicolumn{1}{c}{-}&\multicolumn{1}{c}{-}&\multicolumn{1}{c}{-}&\multicolumn{1}{c}{-}\\
NGC3198&\num{43.354085}&\num{157}&\num{2}&\multicolumn{1}{c}{-}&\multicolumn{1}{c}{-}&\multicolumn{1}{c}{-}&\multicolumn{1}{c}{-}&\multicolumn{1}{c}{-}&\multicolumn{1}{c}{-}&\multicolumn{1}{c}{-}&\multicolumn{1}{c}{-}\\
NGC3198&\num{49.587199}&\num{153}&\num{2}&\multicolumn{1}{c}{-}&\multicolumn{1}{c}{-}&\multicolumn{1}{c}{-}&\multicolumn{1}{c}{-}&\multicolumn{1}{c}{-}&\multicolumn{1}{c}{-}&\multicolumn{1}{c}{-}&\multicolumn{1}{c}{-}\\
NGC3198&\num{55.943741}&\num{153}&\num{2}&\multicolumn{1}{c}{-}&\multicolumn{1}{c}{-}&\multicolumn{1}{c}{-}&\multicolumn{1}{c}{-}&\multicolumn{1}{c}{-}&\multicolumn{1}{c}{-}&\multicolumn{1}{c}{-}&\multicolumn{1}{c}{-}\\
NGC3198&\num{61.868285}&\num{154}&\num{2}&\multicolumn{1}{c}{-}&\multicolumn{1}{c}{-}&\multicolumn{1}{c}{-}&\multicolumn{1}{c}{-}&\multicolumn{1}{c}{-}&\multicolumn{1}{c}{-}&\multicolumn{1}{c}{-}&\multicolumn{1}{c}{-}\\
NGC3198&\num{68.255684}&\num{153}&\num{2}&\multicolumn{1}{c}{-}&\multicolumn{1}{c}{-}&\multicolumn{1}{c}{-}&\multicolumn{1}{c}{-}&\multicolumn{1}{c}{-}&\multicolumn{1}{c}{-}&\multicolumn{1}{c}{-}&\multicolumn{1}{c}{-}\\
NGC3198&\num{74.149371}&\num{150}&\num{2}&\multicolumn{1}{c}{-}&\multicolumn{1}{c}{-}&\multicolumn{1}{c}{-}&\multicolumn{1}{c}{-}&\multicolumn{1}{c}{-}&\multicolumn{1}{c}{-}&\multicolumn{1}{c}{-}&\multicolumn{1}{c}{-}\\
NGC3198&\num{80.53677}&\num{149}&\num{2}&\multicolumn{1}{c}{-}&\multicolumn{1}{c}{-}&\multicolumn{1}{c}{-}&\multicolumn{1}{c}{-}&\multicolumn{1}{c}{-}&\multicolumn{1}{c}{-}&\multicolumn{1}{c}{-}&\multicolumn{1}{c}{-}\\
NGC3198&\num{86.893312}&\num{148}&\num{2}&\multicolumn{1}{c}{-}&\multicolumn{1}{c}{-}&\multicolumn{1}{c}{-}&\multicolumn{1}{c}{-}&\multicolumn{1}{c}{-}&\multicolumn{1}{c}{-}&\multicolumn{1}{c}{-}&\multicolumn{1}{c}{-}\\
NGC3198&\num{92.817856}&\num{146}&\num{2}&\multicolumn{1}{c}{-}&\multicolumn{1}{c}{-}&\multicolumn{1}{c}{-}&\multicolumn{1}{c}{-}&\multicolumn{1}{c}{-}&\multicolumn{1}{c}{-}&\multicolumn{1}{c}{-}&\multicolumn{1}{c}{-}\\
NGC3198&\num{99.174398}&\num{147}&\num{2}&\multicolumn{1}{c}{-}&\multicolumn{1}{c}{-}&\multicolumn{1}{c}{-}&\multicolumn{1}{c}{-}&\multicolumn{1}{c}{-}&\multicolumn{1}{c}{-}&\multicolumn{1}{c}{-}&\multicolumn{1}{c}{-}\\
NGC3198&\num{105.098942}&\num{148}&\num{2}&\multicolumn{1}{c}{-}&\multicolumn{1}{c}{-}&\multicolumn{1}{c}{-}&\multicolumn{1}{c}{-}&\multicolumn{1}{c}{-}&\multicolumn{1}{c}{-}&\multicolumn{1}{c}{-}&\multicolumn{1}{c}{-}\\
NGC3198&\num{111.455484}&\num{148}&\num{2}&\multicolumn{1}{c}{-}&\multicolumn{1}{c}{-}&\multicolumn{1}{c}{-}&\multicolumn{1}{c}{-}&\multicolumn{1}{c}{-}&\multicolumn{1}{c}{-}&\multicolumn{1}{c}{-}&\multicolumn{1}{c}{-}\\
NGC3198&\num{117.842883}&\num{149}&\num{2}&\multicolumn{1}{c}{-}&\multicolumn{1}{c}{-}&\multicolumn{1}{c}{-}&\multicolumn{1}{c}{-}&\multicolumn{1}{c}{-}&\multicolumn{1}{c}{-}&\multicolumn{1}{c}{-}&\multicolumn{1}{c}{-}\\
NGC3198&\num{123.73657}&\num{150}&\num{2}&\multicolumn{1}{c}{-}&\multicolumn{1}{c}{-}&\multicolumn{1}{c}{-}&\multicolumn{1}{c}{-}&\multicolumn{1}{c}{-}&\multicolumn{1}{c}{-}&\multicolumn{1}{c}{-}&\multicolumn{1}{c}{-}\\
NGC3198&\num{130.123969}&\num{150}&\num{3}&\multicolumn{1}{c}{-}&\multicolumn{1}{c}{-}&\multicolumn{1}{c}{-}&\multicolumn{1}{c}{-}&\multicolumn{1}{c}{-}&\multicolumn{1}{c}{-}&\multicolumn{1}{c}{-}&\multicolumn{1}{c}{-}\\
NGC3198&\num{136.017656}&\num{149}&\num{3}&\multicolumn{1}{c}{-}&\multicolumn{1}{c}{-}&\multicolumn{1}{c}{-}&\multicolumn{1}{c}{-}&\multicolumn{1}{c}{-}&\multicolumn{1}{c}{-}&\multicolumn{1}{c}{-}&\multicolumn{1}{c}{-}\\
\hline
\end{tabular}
\end{center}
\end{table}

\newpage
\twocolumngrid
\normalsize

\section{VI. Conclusion}
In this paper we have considered galactic rotation curves in $f(T)$ gravity while taking $f(T)=T+\alpha{T^n}$ as the working model of the gravitational action. A weak field metric \cite{Matteo2015} was utilized for this purpose. We compared various regions of values for the index $n$ with observations of the Milky Way, and three other galaxies which represented other galaxy morphologies. In all cases, the galactic velocity profile was dealt with in terms of a bulge and a disk radial region. This way of decomposing a galaxy better explains the various segments that make up the profile.  For every fitted parameter being considered, Eq.(\ref{eq.all_vect_vel}) was used to determine the combined effect on the final velocity profile. \medskip

The Milky Way galaxy was then used to determine which regions of $n$ are more realistic than others. For each index value, we fit the unknown coupling parameter $\alpha$. Besides the effect on the value of this constant, the various values of $n$ also had an effect on the units of the constant. In the grid figures of Fig.(\ref{fig.grid}), we show plots of all the regions under consideration. The index regions of negative $n$, $n=0$, and $0<n<1$ are discarded due to their velocity profile never vanishing for very large radii. Moreover, the intermediate region behaves very poorly in several instances of those regions. \medskip

The most promising region for $n$ is $1<n<\frac{3}{2}$. For values of $n$ close to $1$, rotation curves were produced which fit very well with large values of $R$. These curves also harbored the possibility of eventually going to zero if the limitations of the functions used were to be surpassed. Such a property is also important as it means that galactic potentials would not interfere on the cosmological scale which would cause serious problems in the theory. The only disadvantage of this range of values of $n$ with this type of fitting is that there is a slight overshooting of the velocity profile for the region of $R$ between $0.1\times{10^{18}}km$ and $0.4\times{10^{18}}km$. That being said, the curve is still within the error bars for a large part of the data points available. One potential solution would be to consider a variation of the teleparallel Lagrangian being investigated in this work, possibly a combination of power laws. Unfortunately, as of yet metrics for such Lagrangians do not exist in teleparallel gravity but are intended to be developed and tested in the near future.  \medskip

For the Milky Way, the best fit galactic velocity profile in this formulation of $f(T)$ gravity was found when applying a multi parameter fitting instead of just a fitting for the coupling constant $\alpha$. Through this fitting we conclude that $n=1.0001$ gives the best fit accompanied by a bulge surface mass density of $5.94\times10^6\;\text{kg}\;\text{km}^{-2}$, a disk mass of $4.08\times10^{10}M_\odot$ and a coupling constant of \num{-4.50573152405E-34}$\;\text{km}^{2n}\;\text{kg}^{-1}\;\text{s}^{-2}$. With these parameters no problems are observed in the small $R$ limit and it fits perfectly with the whole data range given while conserving the possibility that the curve eventually falls to zero. The resulting masses for the galactic components were also consistent with current luminous matter estimates.\medskip

The other three galaxies that were considered were NGC 3198 (bright spiral galaxy \cite{NGCMass}), UGC 128 (low surface brightness disk galaxy \cite{UGC128desk}), and DDO 154 (an irregular dwarf galaxy). These were considered to test the dexterity of the general result for large variance in the components being considered. In Figs.(\ref{fig.grid3}.1-\ref{fig.grid3}.2), a bright spiral galaxy and a low surface brightness disk galaxy are fit consistently for parameter values also found for the Milky Way. This shows the broadness of the result. In Ref.\cite{Capozziello:2006ph} a series of low surface brightness galaxies were considered in the power-law incarnation of the generalized $f(R)$ class of theories. The results in that case were similarly relatively good when compared to GR. These types of galaxies are supposed to be dark matter dominated in comparison to other types of galaxies however using power law models in either $f(R)$ or $f(T)$ (present case) the rotation curve profiles can be almost entirely accounted for using a modified gravitational action.  \medskip

In the case of the irregular galaxy shown in Fig.(\ref{fig.grid3}.3), the fit is very poor due to the dominance of a third component in the galaxy dynamics, namely the role of gas. It would be very interesting to model this component using the $f(T)$ Lagrangian being advanced in this work. In Ref.\cite{Capozziello:2017rvz}, this very case is considered for the power-law model in $f(R)$ theory. The results are very promising and cast further doubt on the need for a dark matter contribution to describe the galactic rotation curve profile. This would be the natural next step for our analysis. \medskip

While the work conducted is promising, it is of paramount importance that such results are tested with regards to larger galaxies surveys that include weighted proportions of the various galaxy morphology types. The parameters obtained here are to be compared with those obtained in each case and if the model allows, a best fit for all galaxy types is to be acquired. This proposed analysis would also test the universality of the coupling parameters, namely $n$ and $\alpha$. A larger multi-galaxy analysis of the results obtained here is thus intended to be conducted in a future work. Moreover, it would be of utmost interest to compare this model with other modified theories of gravity, such as the $f(R)$ gravity models advanced in Refs.\cite{Capozziello:2006ph,Capozziello:2017rvz} among others. Beyond the question of the need for dark matter, the galactic rotation curve profile problem may also advance the question of a preferred model of gravity. \medskip

\section{Acknowledgments}
A. Finch wishes to thank the Institute of Space Sciences and Astronomy at the University of Malta for their support and Dr Joseph Caruana for his valuable comments on the manuscript. The research work disclosed in this publication is partially funded by the Endeavour Scholarship Scheme (Malta). Scholarships are part-financed by the European Union - European Social Fund (ESF) - Operational Programme II - Cohesion Policy 2014-2020 `Investing in human capital to create more opportunities and promote the well-being of society'.

\bibliographystyle{plain}

\begin{thebibliography}{10}

\bibitem{sanders2010dark}
R.H. Sanders.
\newblock {\em The Dark Matter Problem: A Historical Perspective}.
\newblock Cambridge University Press, 2010.

\bibitem{1538-3873-112-772-747}
Vera~C. Rubin.
\newblock One hundred years of rotating galaxies.
\newblock {\em Publications of the Astronomical Society of the Pacific},
  112(772):747, 2000.

\bibitem{Bertone:2004pz}
Gianfranco Bertone, Dan Hooper, and Joseph Silk.
\newblock {Particle dark matter: Evidence, candidates and constraints}.
\newblock {\em Phys. Rept.}, 405:279--390, 2005.

\bibitem{Bi:2014hpa}
Xiao-Jun Bi, Peng-Fei Yin, and Qiang Yuan.
\newblock {Status of Dark Matter Detection}.
\newblock {\em Front. Phys.(Beijing)}, 8:794--827, 2013.

\bibitem{Cirelli:2012tf}
Marco Cirelli.
\newblock {Indirect Searches for Dark Matter: a status review}.
\newblock {\em Pramana}, 79:1021--1043, 2012.

\bibitem{Freeman1970}
K.~C. {Freeman}.
\newblock {On the Disks of Spiral and S0 Galaxies}.
\newblock {\em \apj}, 160:811, June 1970.

\bibitem{McGaugh:2014xfa}
Stacy~S. McGaugh.
\newblock {The Third Law of Galactic Rotation}.
\newblock {\em Galaxies}, 2(4):601--622, 2014.

\bibitem{watson1995treatise}
G.N. Watson.
\newblock {\em A Treatise on the Theory of Bessel Functions}.
\newblock Cambridge Mathematical Library. Cambridge University Press, 1995.

\bibitem{Baojiu}
{Li} B, {Sotiriou} TP, {Barrow} JD.
\newblock {f(T) gravity and local Lorentz invariance}.
\newblock prd. 2011 Mar;83(6):064035.

\bibitem{Sofue:2008wt}
Y.~Sofue, M.~Honma, and T.~Omodaka.
\newblock {Unified Rotation Curve of the Galaxy -- Decomposition into de
  Vaucouleurs Bulge, Disk, Dark Halo, and the 9-kpc Rotation Dip --}.
\newblock {\em Publ. Astron. Soc. Jap.}, 61:227, 2009.

\bibitem{1987gady.book.....B}
J.~{Binney} and S.~{Tremaine}.
\newblock {\em {Galactic dynamics}}.
\newblock 1987.

\bibitem{einstein1923}
A.~{Einstein}.
\newblock {\em {Hamiltonisches prinzip und allgemeine relativitatstheorie Das Relativitatsprinzip}}.
\newblock Berlin: Verlag der Akademie der Wissenschaften, 1923.

\bibitem{Jamil:2012ju}
Mubasher Jamil, D.~Momeni, and Ratbay Myrzakulov.
\newblock {Resolution of dark matter problem in f(T) gravity}.
\newblock {\em Eur. Phys. J.}, C72:2122, 2012.

\bibitem{Krssak:2015oua}
Martin Kr\v{s}\v{s}\'{a}k and Emmanuel~N. Saridakis.
\newblock {The covariant formulation of f(T) gravity}.
\newblock {\em Class. Quant. Grav.}, 33(11):115009, 2016.

\bibitem{Saridakis2016}
Yi-Fu Cai, Salvatore Capozziello, Mariafelicia De~Laurentis, and Emmanuel~N.
  Saridakis.
\newblock {f(T) teleparallel gravity and cosmology}.
\newblock {\em Rept. Prog. Phys.}, 79(10):106901, 2016.

\bibitem{aldrovandi2012teleparallel}
R.~Aldrovandi and J.G. Pereira.
\newblock {\em Teleparallel Gravity: An Introduction}.
\newblock Fundamental Theories of Physics. Springer Netherlands, 2012.

\bibitem{Tamanini:2012hg}
Nicola Tamanini and Christian~G. Boehmer.
\newblock {Good and bad tetrads in f(T) gravity}.
\newblock {\em Phys. Rev.}, D86:044009, 2012.

\bibitem{Hayashi:1979qx}
Kenji Hayashi and Takeshi Shirafuji.
\newblock {New General Relativity}.
\newblock {\em Phys. Rev.}, D19:3524--3553, 1979.
\newblock [Addendum: Phys. Rev.D24,3312(1981)].

\bibitem{Bahamonde:2015zma}
Sebastian Bahamonde, Christian~G. Bohmer, and Matthew Wright.
\newblock {Modified teleparallel theories of gravity}.
\newblock {\em Phys. Rev.}, D92(10):104042, 2015.

\bibitem{Sotiriou:2008rp}
Thomas~P. Sotiriou and Valerio Faraoni.
\newblock {f(R) Theories Of Gravity}.
\newblock {\em Rev. Mod. Phys.}, 82:451--497, 2010.

\bibitem{Capozziello:2011et}
Salvatore Capozziello and Mariafelicia De~Laurentis.
\newblock {Extended Theories of Gravity}.
\newblock {\em Phys. Rept.}, 509:167--321, 2011.

\bibitem{Bahamonde:2016cul}
Sebastian Bahamonde, M.~Zubair, and G.~Abbas.
\newblock {Thermodynamics and cosmological reconstruction in $f(T,B)$ gravity}.
\newblock 2016.

\bibitem{Matteo2015}
M.~L. {Ruggiero} and N.~{Radicella}.
\newblock {Weak-field spherically symmetric solutions in f (T ) gravity}.
\newblock {\em \prd}, 91(10):104014, May 2015.

\bibitem{wald1984general}
R.M. Wald.
\newblock {\em General Relativity}.
\newblock University of Chicago Press, 1984.

\bibitem{Toomre1963}
A.~{Toomre}.
\newblock {On the Distribution of Matter Within Highly Flattened Galaxies.}
\newblock {\em \apj}, 138:385, August 1963.

\bibitem{1983MNRAS.203..735C}
S.~{Casertano}.
\newblock {Rotation curve of the edge-on spiral galaxy NGC 5907: disc and halo
  masses}.
\newblock {\em Mon. Not. R. Astron. Soc.}, 203:735--747, May 1983.

\bibitem{Milgrom:1983pn}
M.~Milgrom.
\newblock {A Modification of the Newtonian dynamics: Implications for
  galaxies}.
\newblock {\em Astrophys. J.}, 270:371--383, 1983.

\bibitem{Mannheim2006}
P.~D. {Mannheim}.
\newblock {Alternatives to dark matter and dark energy}.
\newblock {\em Progress in Particle and Nuclear Physics}, 56:340--445, April
  2006.

\bibitem{Sofue2013}
Y.~{Sofue}.
\newblock {\em {Mass Distribution and Rotation Curve in the Galaxy}}, page 985.
\newblock 2013.

\bibitem{Sofue2008}
Y.~{Sofue}, M.~{Honma}, and T.~{Omodaka}.
\newblock {Unified Rotation Curve of the Galaxy -- Decomposition into de
  Vaucouleurs Bulge, Disk, Dark Halo, and the 9-kpc Rotation Dip --}.
\newblock {\em PASJ}, 61:227--236, February 2009.

\bibitem{andrews1985special}
L.C. Andrews.
\newblock {\em Special Functions for Engineers and Applied Mathematicians}.
\newblock Macmillan, 1985.

\bibitem{Data2}
Pijushpani Bhattacharjee, Soumini Chaudhury, and Susmita Kundu.
\newblock Rotation curve of the milky way out to ~200 kpc.
\newblock {\em The Astrophysical Journal}, 785(1):63, 2014.


\bibitem{SPARC2016}
{{Lelli}, F. and {McGaugh}, S.~S. and {Schombert}, J.~M.}
\newblock {SPARC: Mass Models for 175 Disk Galaxies with Spitzer Photometry and Accurate Rotation Curves}
\newblock {\em The Astronomical Journal}, 152:157, (2016).

\bibitem{1998DDO154}
{Claude Carignan and Chris Purton}
\newblock {The "Total" Mass of DDO 154}
\newblock {\em The Astronomical Journal}, 506(1):125, (1998).

\bibitem{Mass1}
{T.~C. {Licquia} and J.~A. {Newman}}
\newblock {Improved Estimates of the Milky Way's Stellar Mass and Star Formation Rate from Hierarchical Bayesian Meta-Analysis}
\newblock {\em The Astronomical Journal}, 806:96, (2015).

\bibitem{Mass2}
{G.~M. {Eadie} and W.~E. {Harris}, }
\newblock {Bayesian Mass Estimates of the Milky Way: The Dark and Light Sides of Parameter Assumptions}
\newblock {\em The Astronomical Journal}, 829:108, (2016).

\bibitem{NGCMass}
{E V Karukes and P Salucci}
\newblock {Modeling the Mass Distribution in the Spiral Galaxy NGC 3198}
\newblock {\em Journal of Physics: Conference Series}, 566(1):012008, (2014).

\bibitem{UGCMass}
{G. Hensler, G. Stasinska, S. Harfst, P. Kroupa and C.Theis.}
\newblock {The Evolution of Galaxies: III - From Simple Approaches to Self-Consistent Models}
\newblock {\em Springer Netherlands}, 978-94-017-3315-1, (2003).

\bibitem{DDOMass}
{Claude Carignan and Chris Purton}
\newblock {The "Total" Mass of DDO 154}
\newblock {\em The Astrophysical Journal}, 506(1):125, (1998).

\bibitem{UGC128desk}
{Verheijen Marc and de Blok Erwin}
\newblock {The HSB/LSB Galaxies NGC 2403 and UGC 128}
\newblock {\em Astrophysics and Space Science}, 269:673--674, (1999).

\bibitem{Capozziello:2006ph} 
  S.~Capozziello, V.~F.~Cardone and A.~-Troisi,
  Mon.\ Not.\ Roy.\ Astron.\ Soc.\  {\bf 375}, 1423 (2007)
  doi:10.1111/j.1365-2966.2007.11401.x
  [astro-ph/0603522].
  
\bibitem{Capozziello:2017rvz} 
  S.~Capozziello, P.~Jovanović, V.~B.~Jovanović and D.~Borka,
  JCAP {\bf 1706}, no. 06, 044 (2017)
  doi:10.1088/1475-7516/2017/06/044
  [arXiv:1702.03430 [gr-qc]].

\end{thebibliography}

\appendix

\section{Appendix I: Calculating the velocity profile for the range $0 < n < 1$}
For the region $0 < n < 1$, the velocity curve can be represented through the integral
\begin{align}\label{eq. SA1}
V_{e_{{}_{\alpha{d}}}}&=-\dfrac{{\alpha} 2^{3n-2}M_0N}{(2n-3)\beta_d^22\pi}{\displaystyle\int_0^{\infty}}dR'{\displaystyle\int_0^{2\pi}}d\phi'\nonumber\\
&\times{\displaystyle\int_{-\infty}^{\infty}}dz'R'\delta(z')e^{-\frac{R'}{\beta_d}}\;\;r^{2-2n}.
\end{align}

\noindent Noting that
\begin{align}
r=({R'}^2+{R}^2-2RR'{\cos}&{(\phi-\phi')}\nonumber\\
&+(z-z')^2)^\frac{1}{2},
\end{align}

\noindent we define a new variable $x$ such that  
\begin{align}
x^{1-n}&\equiv r^{(2-2n)}\nonumber\\
&={\left(r^2\right)}^{1-n}\nonumber\\
&=({R'}^2+{R}^2-2RR'{\cos}{(\phi-\phi')}\nonumber\\
&+(z-z')^2)^{1-n}.
\end{align}

Assuming that the value of $n$ is restricted between but not including $0$ and $1$, we define the following inverse transformation  
\begin{align}
F(k)&={\displaystyle\int_0^{\infty}}\dfrac{e^{-xk}}{x}\left(x^{1-n}\right)dx\nonumber\\
&=k^{n-1}\Gamma(1-n),
\end{align}
for the function
\begin{align}
f(x)&=x^{1-n}\nonumber\\
&=\dfrac{1}{\Gamma(n)}{\displaystyle\int_0^{\infty}}xe^{-kx}k^{n-1}dk.
\end{align}

With this new expression for $x^{1-n}$ in hand, we substitute it into Eq.(\ref{eq. SA1}). Assuming a flat disk and noting that 
\begin{equation}\label{eq. zint1}
{\displaystyle\int_{-\infty}^{\infty}}\delta(z')z'e^{-kz'}dz'=0,
\end{equation} 
and 
\begin{equation}\label{eq. zint2}
{\displaystyle\int_{-\infty}^{\infty}}\delta(z')e^{-kz'}dz'=1,
\end{equation} 
we obtain the following form for this equation
\begin{align}
V_{e_{{}_{\alpha{d}}}}&=-\dfrac{{\alpha} 2^{3n-2}M_0N}{(2n-3)\Gamma(n)\beta_d^22\pi}{\displaystyle\int_0^{\infty}}dk\, k^{n-1}\nonumber\\
&\times{\displaystyle\int_0^{\infty}}dR'R'e^{-\frac{R'}{\beta_d}}{\displaystyle\int_0^{2\pi}}d\phi'xe^{-kx}.
\end{align}

Substituting for $x$ and integrating with respect to $\phi'$ we find that  
\begin{align}
{\displaystyle\int_0^{2\pi}}&d\phi'xe^{-kx}=\nonumber\\
&e^{-R^2k-{R'}^2k}\Big[2\pi\left(R^2+{R'}^2\right)I_0\left(2RR'k\right)\nonumber\\
&-4\pi{R'}RI_1\left(2RR'k\right)\Big].
\end{align}

Integrating with respect to $k$ we obtain the following form of the potential 
\begin{align}
{V_{e_{{}_{\alpha{d}}}}}=&-\dfrac{\alpha\;{2^{3n-2}}\;N\;M_0}{(2n-3)\;2\pi\;\beta_d^2}{\displaystyle\int_0^{\infty}}dR'R'e^{-\frac{R'}{{\beta}d}}\nonumber\\
&\left(2\pi\;(R^{2}+R'^{2})^{-1-n}\Gamma(n)\left\{(R^{2}+R'^{2})^{2}\right.\right.\nonumber\\
&\left.\times{}_2F_1\left[\{\frac{n}{2},\frac{1+n}{2}\},\{1\},\frac{4R'^2R^2}{(R^{2}+R'^{2})^{2}}\right]\right\}\nonumber\\
&\left.-2nR'^2R^2{}_2F_1\left[\{\frac{1+n}{2},\frac{2+n}{2}\},\{2\},\right.\right.\\
&\left.\left.\frac{4R'^2R^2}{(R^{2}+R'^{2})^{2}}\right]\right).
\end{align}
this last integral has to be worked through numerically since no analytic solution exists. This is shown in the respective velocity profiles shown in Fig.(\ref{fig.grid}).

\section{Appendix II: Calculating the velocity profile for the range $1 < n < \frac{3}{2}$}
For the case of Lagrangian index in the range $1 < n < \frac{3}{2}$, the velocity curve profile takes on the following effective potential
\begin{align}
V_{e_{{}_{\alpha{d}}}}&=-\dfrac{{\alpha} 2^{3n-2}M_0N}{(2n-3)\beta_d^22\pi}{\displaystyle\int_0^{\infty}}dR'{\displaystyle\int_0^{2\pi}}d\phi'\nonumber\\
&\times{\displaystyle\int_{-\infty}^{\infty}}dz'R'\delta(z')e^{-\frac{R'}{\beta_d}}\;\;r^{2-2n}.
\end{align}
As in the first appendix, we set the following transformation and inverse transformation pair s
\begin{align}
F(k)&={\displaystyle\int_0^{\infty}}J_0(kr)\left(r^{2-2n}\right)dr\nonumber\\
&=\dfrac{4^{1-n}k^{2n-3}\Gamma(\frac{3}{2}-n)}{\Gamma(n-\frac{1}{2})},
\end{align}
and
\begin{align}
f(r)&=r^{2-2n}\nonumber\\
&=\dfrac{2^{3-2n}\Gamma(2-n)}{\Gamma(n-1)}{\displaystyle\int_0^{\infty}}k^{2n-3}J_0(kr)dk.
\end{align} 

Given that the integral involves a Bessel function, we need the following relation to move forward in the calculation \cite{watson1995treatise}
\begin{equation}
J_0(kr)=\sum_{m=-{\infty}}^{\infty}J_m(kR)J_m(kR')e^{mi(\phi-\phi')-k(z-z')},
\end{equation} 

Considering the galaxy as a flat disk and using 
\begin{equation}\label{eq. zint3}
{\displaystyle\int_{-\infty}^{\infty}}\delta(z')e^{kz'}dz'=1,
\end{equation}
we find that
\begin{align}
V_{e_{{}_{\alpha{d}}}}&=-\dfrac{{\alpha} 2^{3n-2}M_0N}{(2n-3)\beta_d^22\pi}\left(\dfrac{2^{3-2n}\Gamma(2-n)}{\Gamma(n-1)}\right)\nonumber\\
&\times{\displaystyle\int_0^{\infty}}dkk^{2n-3}{\displaystyle\int_0^{\infty}}dR'R'e^{-\frac{R'}{\beta_d}}\nonumber\\
&\times{\displaystyle\int_0^{2\pi}}d\phi'\sum_{m=-{\infty}}^{\infty}J_m(kR)J_m(kR')e^{mi(\phi-\phi')}.
\end{align}

We now note that $m=0$ is the only value for which the integral gives a non-zero value for the $\phi'$ integral. After integrating over $R'$  and $k$ as in Ref.\cite{Mannheim2006, watson1995treatise} we obtain the final form of the effective potential contribution
\begin{align}
V_{e_{{}_{\alpha{d}}}}&=-\dfrac{{\alpha} 2^{3n-2}M_0N\Gamma(4-2n)\Gamma(n-\frac{3}{2})}{(2n-3)\beta_d^364}\nonumber\\
&\times\left(\dfrac{4^n\sqrt{\pi}R^{5-2n}{\;}_1F_2\left[\left\{\frac{3}{2}\right\},\left\{\frac{7}{2}-n,\frac{7}{2}-n\right\},\frac{R^2}{4\beta^2}\right]}{\Gamma(n-1)\Gamma(\frac{7}{2}-n)^2}\right.\nonumber\\
&\left.+\dfrac{64\beta^{5-2n}{\;}_1F_2\left[\left\{1-n\right\},\left\{1,n-\frac{3}{2}\right\},\frac{R^2}{4\beta^2}\right]}{\Gamma(1)\Gamma(n-\frac{3}{2})}\right)
\end{align}\medskip

Finally we use Eq.(\ref{eq.pot1star}) to obtain the velocity curve equation for this region.
\begin{align}
v_{e_{{}_{\alpha{d}}}}^2&=\dfrac{\alpha8^{n-2}M_0NR^2\beta^{-5-2n}\Gamma(4-2n)\Gamma(n-\frac{3}{2})}{(2n-3)\Gamma(n-1)}\nonumber\\
&\left(4^2\sqrt{\pi}R^{3-2n}\beta^{2n}\left[4(5-2n)\beta^2\right.\right.\nonumber\\
&\times\frac{{\;}_1F_2\left[\left\{\frac{3}{2}\right\},\left\{\frac{7}{2}-n,\frac{7}{2}-n\right\},\frac{R^2}{4\beta^2}\right]}{\Gamma(\frac{7}{2}-n)^2}\nonumber\\
&\left.+3R^2\frac{{\;}_1F_2\left[\left\{\frac{5}{2}\right\},\left\{\frac{9}{2}-n,\frac{9}{2}-n\right\},\frac{R^2}{4\beta^2}\right]}{\Gamma(\frac{9}{2}-n)^2}\right]\nonumber\\
&\left.-128\beta^5\Gamma(n)\frac{{\;}_1F_2\left[\left\{n\right\},\left\{2,n-\frac{1}{2}\right\},\frac{R^2}{4\beta^2}\right]}{\Gamma(2)\Gamma(n-\frac{1}{2})}\right)
\end{align}

\end{document}